\documentclass[11pt,letterpaper]{article}

\usepackage[margin=1in]{geometry}
\usepackage{setspace}
\usepackage{parskip}

\usepackage[T1]{fontenc}
\usepackage[utf8]{inputenc}
\usepackage{microtype}

\usepackage{amsmath}
\usepackage{amssymb}
\usepackage{amsfonts}
\usepackage{amsthm}
\usepackage{bm}
\numberwithin{equation}{section}
\usepackage{nicefrac}

\newcommand{\C}{{\mathbb C}}

\usepackage{graphicx}
\graphicspath{{./images/}}
\usepackage{float}
\usepackage{booktabs}
\usepackage{multirow}
\usepackage{array}
\usepackage{caption}
\usepackage{subcaption}
\usepackage{algorithm}
\usepackage{algpseudocode}

\usepackage[numbers,sort&compress]{natbib}

\usepackage[hidelinks]{hyperref}
\hypersetup{
    colorlinks=false,
    pdfborder={0 0 0}
}
\usepackage{url}

\usepackage{lineno}

\usepackage{xcolor}
\usepackage{soul}
\usepackage{enumerate}
\usepackage{enumitem}
\setlist{nosep}

\title{\Large\bfseries Neural dynamical systems on ferroelectric compute-in-memory for real-time forecasting}

\author{%
    Keshava Katti$^{1}$, Adithya Selvakumar$^{1}$, \\
    Pratik Chaudhari$^{1,*}$, Deep Jariwala$^{1,*}$
}

\date{}

\begin{document}

\maketitle

\begin{center}
\small
$^{1}$Electrical \& Systems Engineering, University of Pennsylvania, Philadelphia, PA 19104, USA \\
$^{*}$Correspondence: \texttt{pratikac@seas.upenn.edu}, \texttt{dmj@seas.upenn.edu}
\end{center}

\vspace{1em}

\begin{abstract}
\noindent
Neural dynamical systems are expressive temporal predictors that capture continuous-time dynamics through fine-grained state updates. However, this sequential structure maps poorly onto digital hardware optimized for dense matrix operations, a mismatch that analog neuromorphic computing, with its native continuous-time dynamics, can resolve. We introduce FerroNDS, a neuromorphic system built from two analog primitives: an integrator for temporal accumulation and an oscillator for frequency-selective filtering. We map this system onto compute-in-memory hardware based on multi-bit ferrodiodes. A 128-neuron instance of FerroNDS computes short-time Fourier transform and forecasts a 500-ms horizon for periodic, quasi-periodic, and chaotic signals. The system achieves sub-watt real-time operation with per-neuron per-inference energy of 1.64 $\mu$J (200 Hz) and 0.29 $\mu$J (10 kHz), 25-40$\times$ area reduction over SRAM-based digital systems, and per-layer latency of 3.18 ms (200 Hz) and 63.87 $\mu$s (10 kHz). To our knowledge, this is the first end-to-end integration of a ferrodiode into a neuromorphic computational framework, establishing ferroelectric compute-in-memory as a practical substrate for analog neural dynamical systems. 
\end{abstract}

\vspace{0.5em}

\begin{figure}[H]
  \centering
  \includegraphics[width=1.0\textwidth]{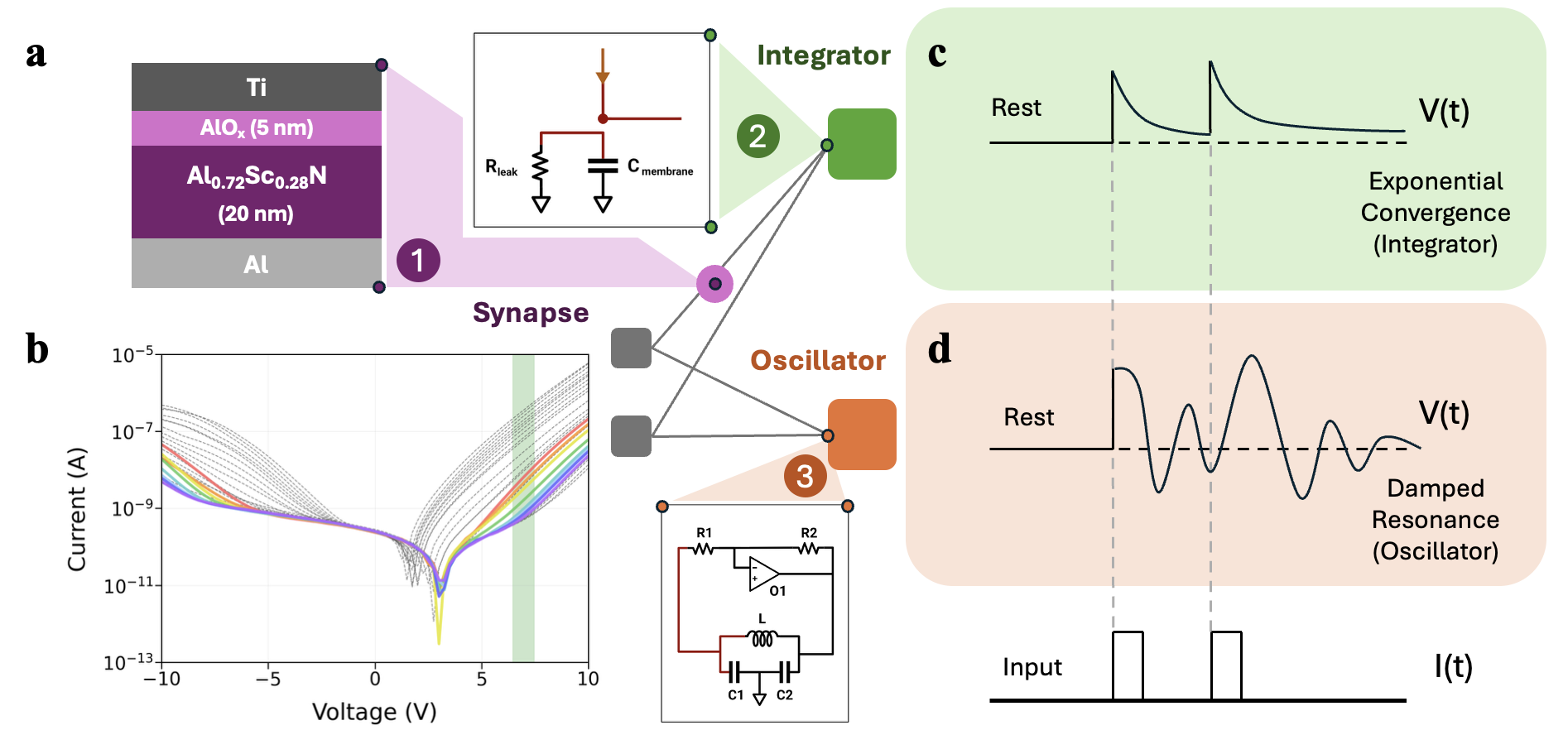}
  \caption{Overview of the FerroNDS system. \textbf{Component 1 (Purple):} (a) CMOS back-end-of-line (BEOL)-compatible ferrodiode (FeD) with 20-nm AlScN film exhibiting up to 5-bit multistate operation and stable retention. (b) 32-state I-V curves of 20-nm AlScN FeD with 5-nm AlO$_x$ interlayer (IL). 8 states (rainbow-colored curves) and corresponding operating regime (shaded green region) selected for maximum exponential behavior \cite{kimMultistateUltrathinBackEndofLineCompatible2024}. \textbf{Component 2 (Green):} (c) Integrator square-pulse response exhibiting exponential convergence with inset showing underlying RC circuit. \textbf{Component 3 (Orange):} (d) Oscillator square-pulse response exhibiting damped resonance with inset showing underlying op-amp and LC-tank circuit \cite{izhikevichResonateandfireNeurons2001}.}
  \label{fig:1}
\end{figure}

\section{Introduction}

Recent work underscores the success of neural dynamical systems (NDS) and state-space models for learning tasks \cite{mehtaNeuralDynamicalSystems2021} \cite{liuKANKolmogorovArnoldNetworks2025} \cite{guEfficientlyModelingLong2022a} \cite{guMambaLinearTimeSequence2024a}. At the heart of these models is a recurrence that advances a latent state one small timestep at a time, capturing continuous-time dynamics with considerable expressivity. Digital accelerators, however, are built around large, dense matrix operations and extract little throughput from sequential updates of this kind, particularly as refresh rates climb into the kilohertz range. Analog circuits sidestep this mismatch because the same differential equations can be realized directly in hardware, with time evolving as a physical quantity rather than a loop index. This alignment has long been recognized in neuromorphic computing, where differential equations are mapped onto networks of leaky integrator neurons and static non-linear functions \cite{eliasmithNeuralEngineeringComputational2002} \cite{neckarBraindropMixedSignalNeuromorphic2019}. Among dynamical primitives, oscillators lend themselves to frequency-based tasks like Fourier analysis and are provably universal approximators of continuous causal operators, while integrators handle temporal accumulation tasks, such as determining position and velocity from linear acceleration \cite{lanthalerNeuralOscillatorsAre2023}. These two operations form the basis of Izhikevich's neuron model, considered one of the most simple and expressive models in computational neuroscience, and they motivate the present work \cite{izhikevichSimpleModelSpiking2003}. 

Analog circuitry is also more efficient at low precision: an analog low-pass filter uses only two physical primitives, a capacitor and a transconductance amplifier, achievable in as few as 5 transistors, whereas a digital counterpart would require 56 logical primitives, each composed of 30 transistors \cite{boahenNeuromorphsProspectus2017}. Realizing these analog dynamics efficiently, however, benefits from memory co-located with compute. Neuromorphic architectures achieve this in one of two forms: compute-near-memory, typically with on-chip SRAM, or compute-in-memory (CIM) using emerging non-volatile memories (eNVMs). Both approaches significantly reduce communication costs, allowing them to consume power on the order of tens or hundreds of milliwatts in contrast to single-board computers, like the NVIDIA Jetson Nano, which are on the order of several watts \cite{hollyProfilingEnergyConsumption2020} \cite{abukhalilPowerOptimizationMobile2020}. Given the favorable energy scaling of CIM, commercial fabrication of eNVMs has been growing but remains limited to resistive random-access memory (RRAM), phase-change memory (PCM), spin-transfer torque magnetic random-access memory (STT-MRAM), and ferroelectric field-effect transistors (FeFETs), all of which offer single-bit storage only \cite{yuComputeinMemoryChipsDeep2021}. Neuromorphic systems with eNVMs are still at an early stage, with demonstrations typically involving neural units in the hundreds, confined initially to inference. Previous work has applied eNVM-based systems to offline learning tasks \cite{yuComputeinMemoryChipsDeep2021} \cite{niFerroelectricTernaryContentaddressable2019} \cite{wanComputeinmemoryChipBased2022} \cite{liu332FullyIntegrated2020}. However, real-time temporal signal analysis is identified as the setting where neuromorphic processors offer the clearest advantage \cite{muirRoadCommercialSuccess2025}.

A growing community of researchers has responded by focusing on dynamic, online tasks, with a parallel thread of device-to-algorithm co-designs in which new eNVMs are mapped to deployable neuromorphic systems through device-level modeling, circuit simulation, and algorithmic design, whether the underlying hardware is fabricated or simulated. NeuroBench consolidates this turn toward dynamic workloads with a set of standardized datasets, including keyword spotting from streamed audio data, event camera object detection, and chaotic function prediction \cite{yikNeurobenchFrameworkBenchmarking2025}. Recent proposals span a wide range of device-algorithm pairings and workload scales. DenRAM leverages a fabricated RRAM memory array for heartbeat anomaly detection and spoken digit classification \cite{dagostinoDenRAMNeuromorphicDendritic2024a}. A three-terminal Li$_x$WO$_3$ memristor has been integrated in simulation into a time-surfaces architecture for event-based tasks such as N-MNIST and POKER-DVS \cite{rasettoBuildingTimeSurfacesExploiting2023}. Simulation-based studies have also mapped the attention mechanism of large language models and evaluated mixture-of-experts architectures on analog in-memory computing \cite{lerouxAnalogInmemoryComputing2025} \cite{buchelEfficientScalingLarge2025}. Collectively, these works demonstrate that end-to-end co-designs, whether realized in silicon or through simulation, are a recognized format for advancing neuromorphic systems.

Among emerging memory devices, multi-bit eNVMs are increasingly of interest to the neuromorphic device community due to their ability to reduce the total device count in a neuromorphic system, as well as interface seamlessly with analog circuits. One such candidate device is a ferrodiode (FeD), a two-terminal ferroelectric CIM device that has demonstrated 5-bit operation, 10 fJ/bit switching energy, ultra-fast switching speed, and as little as a 2 V operating voltage \cite{kimMultistateUltrathinBackEndofLineCompatible2024} \cite{sarkarMultistateFerroelectricDiodes2024} \cite{sarkarCanFerroelectricDiode2025}. Additionally, FeDs utilize the inherent property of ferroelectric polarization-dependent leakage current to form a memory device with built-in rectification, eliminating the need for an additional selector element and in turn driving down the per-cell area in a memory array \cite{huDemonstrationHighlyScaled2025}. Yet no existing work has combined these two ingredients, multi-bit FeDs and analog oscillator/integrator dynamics, into an end-to-end neuromorphic system for a concrete real-time signal processing task.

We introduce FerroNDS, an end-to-end FeD-based neuromorphic system that combines oscillator and integrator dynamical systems for real-time signal prediction. While the underlying framework is device-agnostic, we instantiate it here with an FeD for the reasons above. The contributions of this work are as follows. First, we present a linearizing circuit for the FeD that enables its integration into a synaptic weighting circuit with a stable input-output range of [0, 3.3] V (Fig. \ref{fig:1}(b)), along with an algorithm to identify the optimal subset of FeD states and corresponding operating voltages. Second, we present stable oscillator and integrator circuits that operate up to 200 Hz and 10 kHz (Fig. \ref{fig:1}(c, d)). Third, and most centrally, we show that a 128-neuron FerroNDS computes short-time Fourier transform (STFT) and predicts a 500-ms horizon into the future of complex one-dimensional functions with a mean-squared error (MSE) of $\leq$ 0.32. Fourth, we benchmark the system: per-neuron per-inference energy of 1.64 $\mu$J (200 Hz) and 0.29 $\mu$J (10 kHz), 25-40$\times$ area reduction over SRAM-based digital systems, and per-layer latency of 3.18 ms (200 Hz) and 63.87 $\mu$s (10 kHz). Taken together, these contributions form an eNVM-agnostic device-circuit-algorithm framework for deploying neural dynamical systems in real-time, low-power applications. Circuit schematics are designed and simulated in LTspice; all modeling and prediction code is implemented in Python and included in Methods.

\section{Results}

\subsection{Multi-bit ferrodiode for synaptic weighting} \label{sec:2}

Several ferroelectric memory devices have been explored for compute-in-memory applications, each with distinct tradeoffs. Ferroelectric random-access memory (FeRAM) offers high endurance ($10^{15}$ cycles) but requires destructive readout \cite{sarkarCanFerroelectricDiode2025} \cite{fengFerroelectricFinDiode2024}. Ferroelectric tunnel junctions (FTJs) are structurally simpler but need ultra-thin films for direct tunneling, constraining endurance to around $10^6$ cycles \cite{garciaFerroelectricTunnelJunctions2014} \cite{wenFerroelectricTunnelJunctions2020}. Ferroelectric field-effect transistors (FeFETs) are three-terminal devices that exhibit higher device-to-device variation \cite{fengFerroelectricFinDiode2024} \cite{khanFutureFerroelectricFieldeffect2020}. We focus on the ferrodiode (FeD), whose self-rectifying I-V characteristics suppress sneak-path currents in crossbar arrays, eliminating the need for a separate selector element and reducing cell complexity from 1T1C to 1R \cite{sarkarCanFerroelectricDiode2025} 
\cite{huDemonstrationHighlyScaled2025} \cite{shiResearchProgressSolutions2020}. AlScN-based FeDs achieve ON/OFF ratios of $10^5$, rectification ratios of $10^3$, switching speeds of $<$ 1 ns, and endurance up to $10^9$ cycles \cite{sarkarCanFerroelectricDiode2025} \cite{luoHighlyCMOSCompatible2020a} \cite{liuAluminumScandiumNitridebased2021}.

\begin{figure}[tbp]
  \centering
  \includegraphics[width=1.0\textwidth]{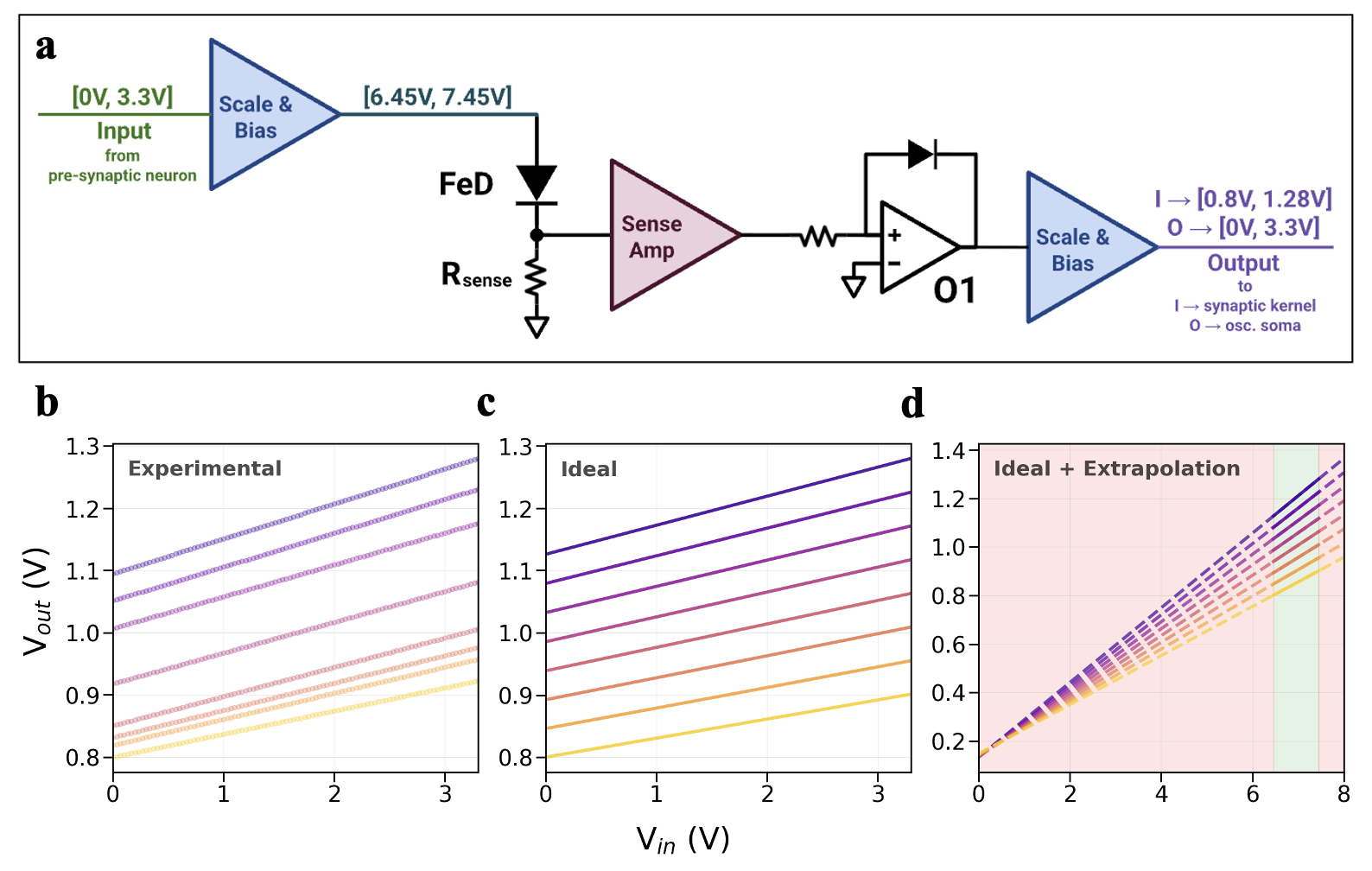}
  \caption{FeD-based synaptic weighting circuit. (a) Proposed five-stage circuit, designed and simulated in LTspice. See ``Linearizing the Synaptic Weight'' for more detail. \textbf{Stage 1:} Scale/bias block that shifts input to selected operation regime of FeD; \textbf{Stage 2:} Transimpedance amplifier block that leverages massive resistance of FeD (modeled as a voltage-dependent current source with input range [6.45, 7.45] V) to convert state-dependent device current back to a voltage without op-amp; \textbf{Stage 3:} Sense amplifier with gain of 1000; \textbf{Stage 4:} Logarithmic amplifier to linearize exponential I-V characteristics of FeD; \textbf{Stage 5:} Scale/bias block that shifts output signal to operation regime of integrator (I) and oscillator (O). (b) Circuit voltage transfer function for 8 selected FeD states exhibiting $A$-coefficient $R^2 = 0.9867$ and $G$-value CV $ = 0.37$. (c) Modeled 8 states in Python with $A$-coefficient $R^2 = 1$ and $G$-value CV $= 0$. (d) Extrapolated modeled states to illustrate $y = m_i x + b_i$ for $i \in \{1, \dots, 8\}$.} 
  \label{fig:2}
\end{figure}

This work focuses on a recently-reported multi-bit FeD with a 20-nm AlScN film and 5-nm AlO$_x$ interlayer (Fig. \ref{fig:1}(a)) \cite{kimMultistateUltrathinBackEndofLineCompatible2024}. A large ON/OFF ratio of 1175 allows for stable maintenance of 32 states (\ref{fig:1}(b)). The main design challenge is that the FeD's transfer characteristics are exponential. Previous work has handled this non-linearity in software via min-max normalization, while the closest circuit-level approach uses a logarithmic amplifier paired with an FTJ \cite{liuReconfigurableComputeInMemoryFieldProgrammable2022} \cite{berdanLowpowerLinearComputation2020}. Our circuit adopts a similar logarithmic motif but replaces the FTJ with an FeD and handles all non-linearity in hardware, requiring no software preprocessing.

\textbf{Linearizing the synaptic weight.} We provide a 5-stage synaptic weighting circuit tailored to an FeD and summarized in Fig. \ref{fig:2}(a). \textit{Stage 1.} A scaling amplifier scales the input from either an external sensor or some arbitrary somatic circuit into the FeD's operating regime $V_{in}$, which is necessary for FerroNDS stability. \textit{Stage 2.} A TIA takes $V_{in}$ as input. The FeD and resistor $R_{sense}$ are placed in series, and $V_{in}$ is held across this series pair. Because the FeD resistance ($\sim 10^9$ $\Omega$) is orders of magnitude larger than $R_{sense}$ ($\sim 10^5$ $\Omega$), effectively all of $V_{in}$ drops across the FeD. As $V_{in}$ changes linearly, the current through the FeD changes exponentially within the operating regime. This current is run through $R_{sense}$, and the resulting voltage $V_{sense}$ becomes the input to the subsequent sense amplifier. Suppose we have a set of $N$ FeD states with fitting coefficients given by $\{(G_i, A_i)\}_{i=1}^N$. The following equation describes the transfer characteristics that relate voltage $V_{in}$ to output current $I_{out}$:
\begin{align}
  I_{out} = G_i \cdot \exp (A_i \cdot V_{in}) \label{2.1}
\end{align}
\textit{Stage 3.} A sense amplifier takes $V_{sense}$, which ranges from $\sim$ 0.3–10 mV, as input. Using a precision instrumentation amplifier with sufficiently low offset voltage, a programmed gain of 1000 reliably amplifies these small voltages to a much larger output $V_{sa}$. \textit{Stage 4.} A logarithmic amplifier produces an output proportional to the natural logarithm of its input and yields the resulting voltage $V_{log}$ given by:
\begin{align}
  V_{log} &= \ln (R_{sense} \cdot G_i \cdot \exp (A_i \cdot V_{sa})) \nonumber \\
  &= \ln (R_{sense} \cdot G_i) + A_i \cdot V_{sa} \label{2.2}
\end{align}
\textit{Stage 5.} An inverting summer takes two inputs: the output of the previous stage $V_{log}$ and a fixed offset voltage $V_{offset}$. Its output $V_{out}$ is scaled to the appropriate input range for the downstream somatic circuit. Both Stage 1 and Stage 5 represent scaling steps that ensure no node in the FerroNDS receives an input or produces an output outside its operating regime. We then rewrite the transfer characteristics in the familiar linear form by substituting $m_i$ for $A_i$, $b_i$ for $\ln(R_{sense} \cdot G_i)$, $x$ for $V_{in}$, and $y$ for $V_{out}$:
\begin{align}
  y = m_i \cdot x + b_i \label{2.3}
\end{align}
For any given state $i \in \{1, \ldots, N\}$.

\textbf{Algorithm for selecting ferrodiode states.} Not all states in a multi-bit eNVM are amenable to a linear weighting circuit. We define four constraints (minimum current level, exponential fit quality, linear spacing of $A_i$ coefficients, and uniformity of $G_i$ coefficients) and search over the 32 I-V curves in Fig. \ref{fig:1}(b) to identify a subset that satisfies all four (details in Methods). The [6.45, 7.45] V range yields 8 states meeting these criteria (Fig. S1). Fig. \ref{fig:2}(b) shows the resulting input-output characteristics of the synaptic weighting circuit.

\textbf{Ideal ferrodiode model.} Based on experimental data corresponding to the 8 selected states, we created a synthetic model that has ideal linearity in the $A_i$ terms (i.e., $R^2 = 1$), as well as ideal uniformity in the $G_i$ terms (i.e., CV $= 0$). These modeled states are shown in Fig. \ref{fig:2}(c). The purpose of this compact model is to provide a target for FeD device designers and a simple framework for exploring the computational capability of FeD-based systems. The resulting states form a set of uniformly-distributed lines originating at the same $y$-intercept, as illustrated in Fig. \ref{fig:2}(d). Given that multi-bit FeDs (with up to 5 states demonstrated) have been shown to operate at 2 V, a future direction of this work is to integrate such devices into FerroNDS such that the green region in Fig. \ref{fig:2}(d) is shifted down to a range that enables lower power consumption (power scales with operating voltage; see Section \ref{sec:5}) \cite{sarkarMultistateFerroelectricDiodes2024}. 

\textbf{Generalization to other multi-bit eNVMs.} The FerroNDS synaptic weighting circuit is not specific to the FeD. Any multi-bit eNVM can be integrated provided: (1) the I-V curves corresponding to each state can be fit to a parametrized function $f$ for which an inverse circuit $f^{-1}$ exists to linearize the transfer characteristics, (2) the device exhibits more states than the target bit-precision, since not all states will satisfy the selection constraints, and (3) scale/bias stages are included to align the eNVM's operating regime with the somatic circuits. The maximum frequency of the op-amps sets the inference rate and dominates power consumption.

\subsection{Integrator and oscillator dynamics} \label{sec:3}

The synaptic weighting circuit is designed to interface with analog somatic circuits, which we turn to next. Our design draws on two primitives from computational neuroscience: the leaky integrate-and-fire (LIF) neuron, which exhibits exponential convergence to rest after each pulse (Fig. \ref{fig:1}(c)), and the resonate-and-fire (RF) neuron, which exhibits damped oscillation to rest (Fig. \ref{fig:1}(d)) \cite{izhikevichDynamicalSystemsNeuroscience2007}. Both are derived formally in Section S3. While spiking implementations of these primitives are well-studied, spiking networks are difficult to train and may require significantly more units than continuous-valued alternatives to achieve comparable performance \cite{pfeifferDeepLearningSpiking2018} \cite{semenovAdvantagesDisadvantagesSpiking2023}. We therefore adopt continuous real-valued integrator and oscillator primitives, accepting higher per-unit power in exchange for greater expressivity at small scale. At the system sizes relevant to FerroNDS, analog computation and communication with multi-bit CIM reduces overall communication overhead \cite{boahenDendrocentricLearningSynthetic2022}. The following two subsections present each primitive as a mathematical model and its underlying analog circuit.

\begin{figure}[tbp]
  \centering
  \includegraphics[width=1.0\textwidth]{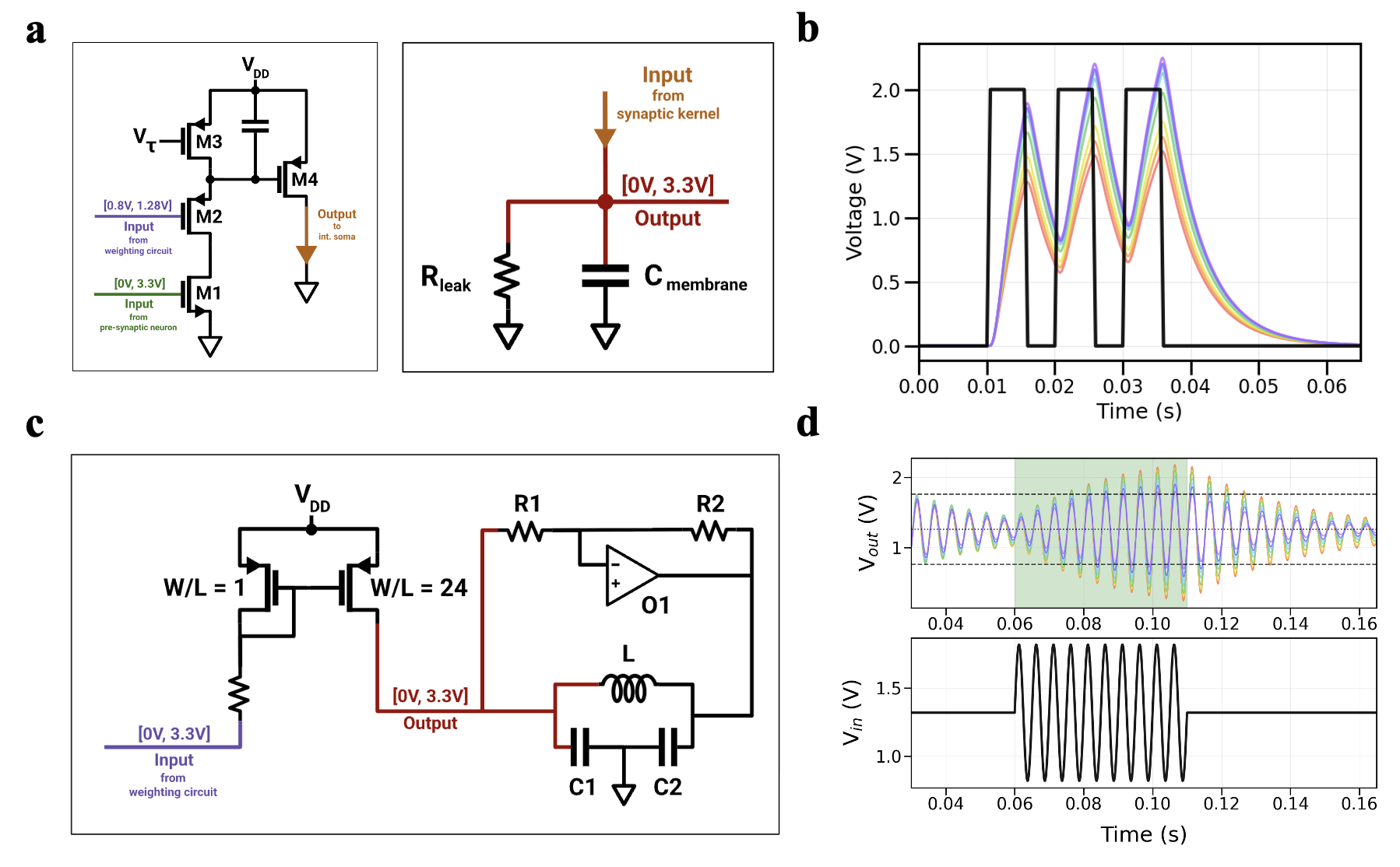}
  \caption{Integrator and oscillator circuits, designed and simulated in LTspice. (1) \textbf{Left:} Charge-discharge module that appends an exponential decay tail to the synaptic response; \textbf{Right:} LI module that sums inputs from the synapse. (b) Response of LI circuit to 3 square pulses across all 8 ferrodiode states. (c) Oscillator module that leverages op-amp feedback and an LC-tank to set damped or sustained oscillation at a specific resonant frequency. (d) Response of BP oscillator circuit to a sine wave at the circuit's resonant frequency, exhibiting a periodic response that grows in magnitude during the applied stimulus and relaxes down via damped oscillation to rest thereafter.}
  \label{fig:3}
\end{figure}

\textbf{Integrator.} We implement the integrator primitive as a standard first-order leaky integrator (LI):
\begin{align}
    \dot{y} = \frac{x - y}{\tau} \label{3.3}
\end{align}
Section S3.1 connects this form of LI to the LIF equation (S1) and derives the discrete implementation:
\begin{align}
    y_k = \alpha y_{k-1} + (1 - \alpha) \phi(x_{k-1}) \label{3.4}
\end{align}
For discrete steps $k \in \mathbb{N}$, $\alpha := \sigma(e^{-\Delta t / \tau})$ and $\phi(x_{k-1}) = f(W x_{k-1})$. Here $\sigma$ denotes an optional sigmoid function that constrains $\alpha \in (0, 1)$ during gradient-based training. $\Delta t$ corresponds to a fixed step size, $\tau$ is a time constant, $f(\cdot)$ can either be the identity function or a non-linearity (e.g., hyperbolic tangent or sigmoid), and $W$ is the synaptic weight matrix. 

We now describe the analog circuit that implements (\ref{3.3}). In addition to our synaptic weighting circuit, we provide a charge-discharge synaptic module, which we also refer to as a synaptic kernel, shown in Fig. \ref{fig:3}(a) (left). This circuit sits before the LI circuit and appends an exponential decay tail to the synaptic response. The exponential decay tail is standard in both computational neuroscience and neuromorphic computing, capturing the time since the last synaptic input \cite{gerstner31SynapsesNeuronal2014} \cite{lagorceHOTSHierarchyEventBased2017}.

Prior to an external stimulus, $V_{CAP}$ charges toward $V_{DD}$ through M3 at a rate set by $V_{\tau}$, leaving near-zero voltage across the capacitor and minimal output. When $V_{IN}$ is applied M1 turns on and current flows through M2, whose gate voltage $V_W$ is set by the synaptic weighting circuit. As $V_{CAP}$ drops below $V_{DD}$, M4 current increases proportionally to $V_{DD} - V_{CAP}$, so each of the 8 FeD states produces a distinct post-synaptic response. After $V_{IN}$ is removed, M1 turns off and $V_{CAP}$ recharges through M3, producing an exponential decay in the M4 output. The complete synapse-to-soma pathway appears in Fig. \ref{fig:3}(a): the charge-discharge synapse (left) drives the somatic membrane circuit (right). Fig. \ref{fig:3}(b) shows the response to three square pulses, with the rainbow-colored curves corresponding to the 8 FeD synaptic states. The somatic membrane voltage $V_{MEM}$ progressively accumulates charge with each input pulse, partially decaying between pulses through $R_{leak}$ but not returning to baseline, enabling temporal integration of successive inputs.

\textbf{Oscillator.} The oscillator model that we use is based on a harmonic resonate-and-fire (HRF) neuron, which describes changes in membrane potential with the dynamics of a damped harmonic oscillator \cite{alkhamissiDeepSpikingNeural2021} \cite{higuchiBalancedResonateandFireNeurons2024}. Rather than using the complex-valued representation seen in (S2), each neuron's latent state $(z_1, z_2)$ is described as:
\begin{align}
    \dot{z_1} &= -2 \xi \omega z_1 - \omega^2 z_2 + x \nonumber \\
    \dot{z_2} &= z_1 \label{3.5}
\end{align}
Where $\omega$ is the neuron's resonant frequency and $\xi$ is the damping coefficient. The dynamical system given by (\ref{3.5}) is a bandpass (BP) oscillator, the term we use going forward. We derive the BP oscillator from the RF neuron in Section S3.2. A discrete version of the BP oscillator is used to model the FerroNDS in Section \ref{sec:4} and can be determined via Euler integration as follows:
\begin{align}
    z_{1, k} &= z_{1, k-1} + \Delta t \left( -2 \xi \omega z_{1, k-1} - \omega^2 z_{2, k-1} + x_k \right) \nonumber \\
    z_{2, k} &= z_{2, k-1} + \Delta t \left(z_{1, k-1} \right) \label{3.6}
\end{align}

We now describe the analog circuit that implements (\ref{3.5}). The circuit shown in Fig. \ref{fig:3}(c) operates as a damped, forced harmonic oscillator. A transistor front-end (with W/L ratio of 1:24) converts the input voltage into a controlled current that sets the forcing amplitude. The oscillator is realized through a feedback loop formed by the op-amp and an LC-tank network, with resonant frequency:
\begin{align}
    f_0 \approx  \frac{1}{2\pi\sqrt{L C_{eq}}} \label{3.7}
\end{align}
Where $C_{eq}$ is the effective capacitance determined by $C_1$, $C_2$, and their connection topology. Component values are chosen to set $f_0 = 200$ Hz, the maximum usable frequency of the op-amp; higher frequencies require higher-bandwidth op-amps at the cost of increased power consumption. The resistive feedback network $(R_1, R_2)$ compensates for the LC-tank losses and shapes the effective damping. When the input contains a frequency component near $f_0$, energy accumulates coherently over successive cycles, producing a large-magnitude response (Fig. \ref{fig:3}(d)); off-resonance inputs produce only a small response (Fig. S1(d)). After the input is removed, the circuit exhibits damped oscillations that decay at a rate set by the feedback gain. In the limiting case of zero damping, oscillations persist indefinitely; we revisit this undamped regime in Section \ref{sec:4}.

\begin{figure}[tbp]
  \centering
  \includegraphics[width=1.0\textwidth]{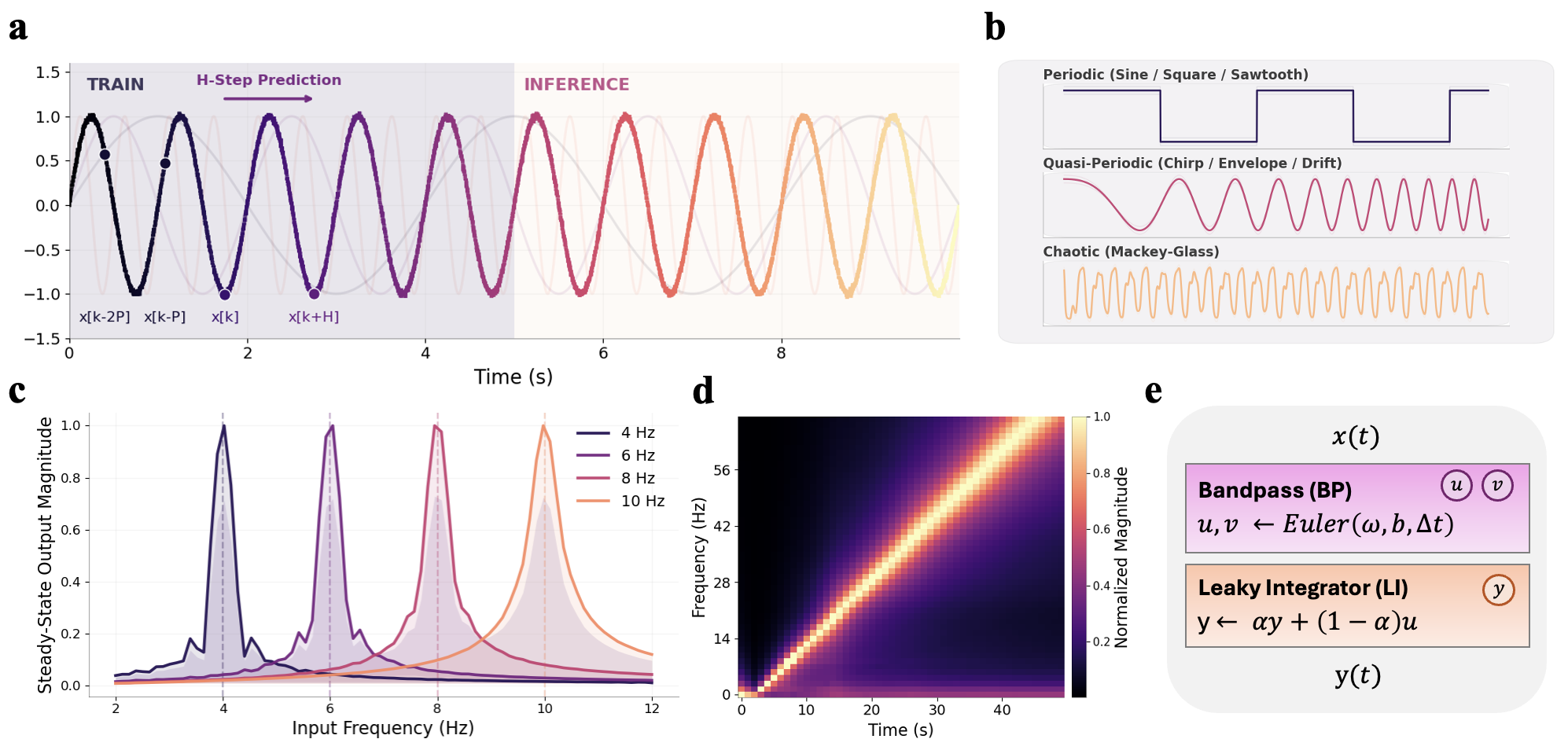}
  \caption{BP/LI Architecture for Horizon Prediction. (a) Illustration of an $H$-ms horizon prediction task. (b) Three main signal classes used to benchmark performance. (c) Frequency response of a set of linearly-spaced BP oscillators shows filter bank behavior. (d) Response to a chirp signal by 50 BP oscillators with resonant frequencies uniformly distributed over 4 to 64 Hz, approximately computing short-time Fourier transform (STFT). (e) BP/LI architecture has a BP block to project input into Fourier-like feature space followed by a LI block that temporally pools these features for horizon prediction.}
  \label{fig:4}
\end{figure}

\subsection{Horizon prediction of signals} \label{sec:4}

We now connect the synaptic weighting circuit from Section \ref{sec:2} with the integrator and oscillator somata from Section \ref{sec:3} to form FerroNDS. We evaluate three architectural instantiations: a feed-forward BP/LI network and a reservoir-style BP pool (both using damped oscillation), plus a UH/LI network (using undamped oscillation to capture longer-range temporal dependencies). We implement the integrator dynamics (\ref{3.4}) in Subroutine S1, damped oscillator dynamics (\ref{3.6}) in Subroutine S2, and undamped dynamics (i.e., (\ref{3.6}) with $\xi = 0$) in Subroutine S4.

\textbf{Horizon prediction task.} We evaluate all three architectures on a horizon prediction task (Methods). Fig. \ref{fig:4}(a) illustrates the task: a signal $x$ is sampled at some discrete values $k \in \mathbb{N}$, and for every $k$, we use the current signal value $x[k]$ to predict the future value $x[k + H]$. As pictured in Fig. \ref{fig:4}(b), we consider three main signal classes: periodic, quasi-periodic, and chaotic. We selected these classes because real-world sensor streams rarely occupy a single dynamical regime. ECG signals exhibit quasi-periodic modulation of a periodic substrate; rotating machinery transitions between steady-state periodicity and transient non-stationarities under varying load; EEG and event camera outputs display chaotic, burst-like temporal structure. A forecasting architecture that learns shared representations across periodic, quasi-periodic, and chaotic generators therefore provides a controlled proxy for the spectral and dynamical heterogeneity encountered across many physical sensing modalities.

\textbf{Damped bandpass (BP).} We see from Fig. \ref{fig:4}(c) that BP oscillators, as their name suggests, are a bank of bandpass filters. This result comes from passing 80 sinusoidal inputs with frequencies ranging across 2-12 Hz into 4 BP oscillators tuned to 4, 6, 8, and 10 Hz and reporting the steady-state response to each input. When a chirp signal ranging from 1-70 Hz over 50 s is fed through 50 BP oscillators linearly-spaced from 4-64 Hz, the collective response approximates the short-time Fourier transform (STFT) of the input (Fig. \ref{fig:4}(d)). This is not a designed property; it emerges from the fact that each oscillator passively integrates energy near its resonant frequency, so a population of oscillators with staggered frequencies implements a time-varying spectral decomposition without any explicit Fourier computation. Analog dynamics produce spectral features as a byproduct of physics. Building on this filter-bank behavior, the BP/LI architecture (Fig. \ref{fig:4}(e); Algorithm S3) consists of 64 BP oscillators with recurrently connected learnable weights projecting the input into a Fourier-like feature space, followed by 64 LI units temporally pooling these features for downstream tasks. The BP layer acts as a learnable adaptive filter bank, while the LI layer integrates its output over time.

\begin{figure}[tbp]
  \centering
  \includegraphics[width=1.0\textwidth]{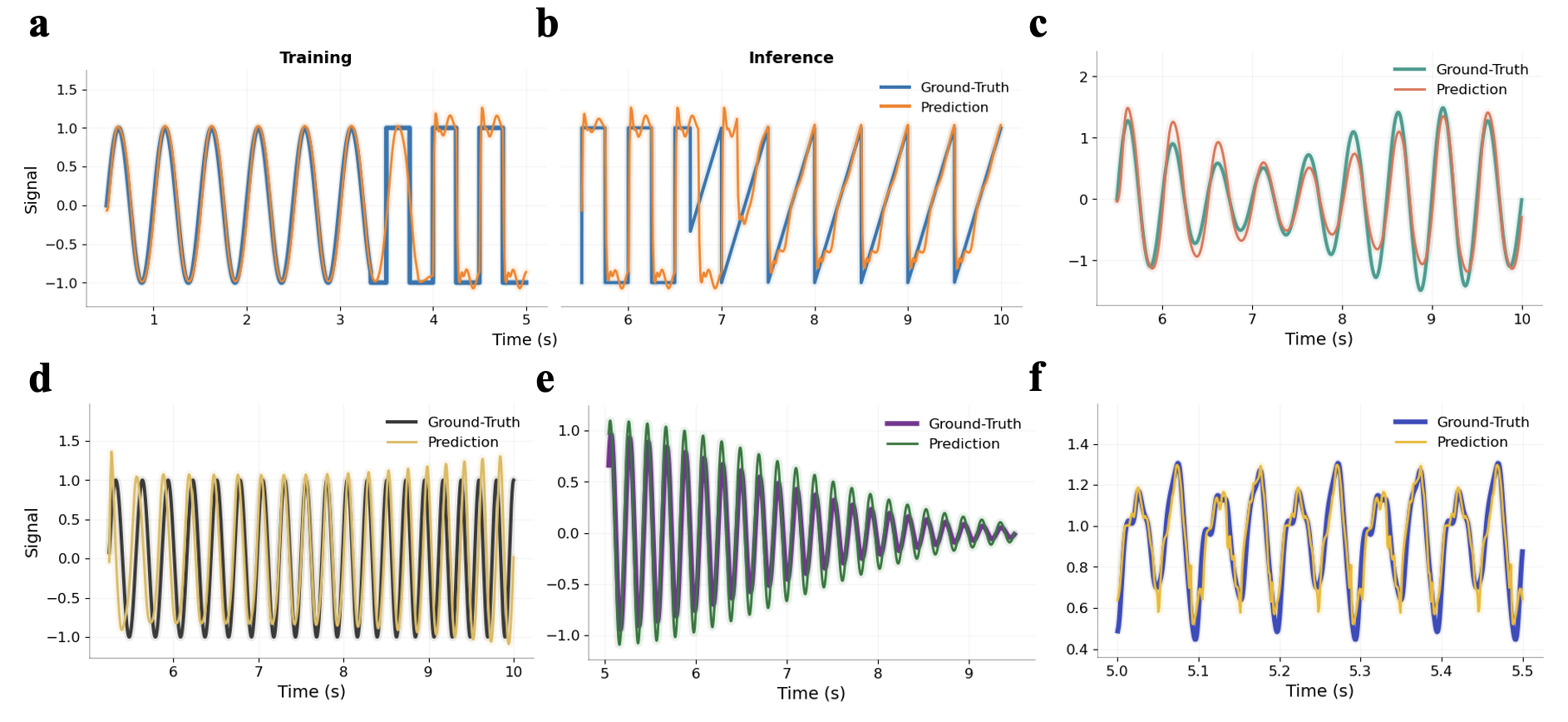}
  \caption{Horizon prediction results. All panels show 500-ms prediction. \textbf{Composite:} (a, b) Piecewise signal consisting of sine and square waves at training time but square and sawtooth waves at inference time. \textbf{AM Sine:} (c) Periodic amplitude-modulated sine function. \textbf{Chirp:} (d) Sine wave with linearly increasing frequency. \textbf{Envelope:} (e) Aperiodic envelope-modulate sine function. \textbf{Mackey-Glass:} (f) Chaotic delay differential equation.}
  \label{fig:5}
\end{figure}

Fig. \ref{fig:5}(a, b) shows a quasi-periodic composite signal constructed to probe out-of-distribution generalization. The composite signal consists of sine, square, and sawtooth segments at 2 Hz. The training data contains only sine and square waves, while inference requires predicting a sawtooth wave that the network never sees during training. We test a reservoir architecture on this task (a fully-connected pool of 128 BP oscillators with a linear readout) and find that it adapts well to the unseen waveform (Fig. \ref{fig:5}(a, b); Table S1). We attribute this to the reservoir's broad spectral basis: a pool of oscillators with diverse, fixed resonant frequencies spans enough frequency content to represent the sawtooth's harmonics, so the readout layer can recombine the oscillator outputs into the sawtooth waveform, even though no sawtooth was seen during training. The feed-forward BP/LI architecture handles periodic amplitude-modulated sine, chirp, and envelope-modulated sine signals at 500-ms horizon (Fig. \ref{fig:5}(c, d, e); Table S1). The chirp is the hardest of these three because its instantaneous frequency changes during inference; the network's filter-bank structure nonetheless tracks the moving frequency content and produces stable predictions. 

As a baseline, we test a standard ReLU-based MLP on the same horizon prediction task (Section S5). The MLP handles periodic signals comfortably but struggles on quasi-periodic alternatives like the chirp and composite signal, where FerroNDS architectures (feed-forward BP/LI on chirp, reservoir BP on composite) reduce inference MSE by roughly 1.6$\times$ (chirp) to 2$\times$ (composite) over the MLP baseline (Table S1). This suggests that architectural alignment matters: a filter-bank representation is better suited to non-stationary periodic structure than a dense feed-forward network.

\begin{figure}[tbp]
  \centering
  \includegraphics[width=1.0\textwidth]{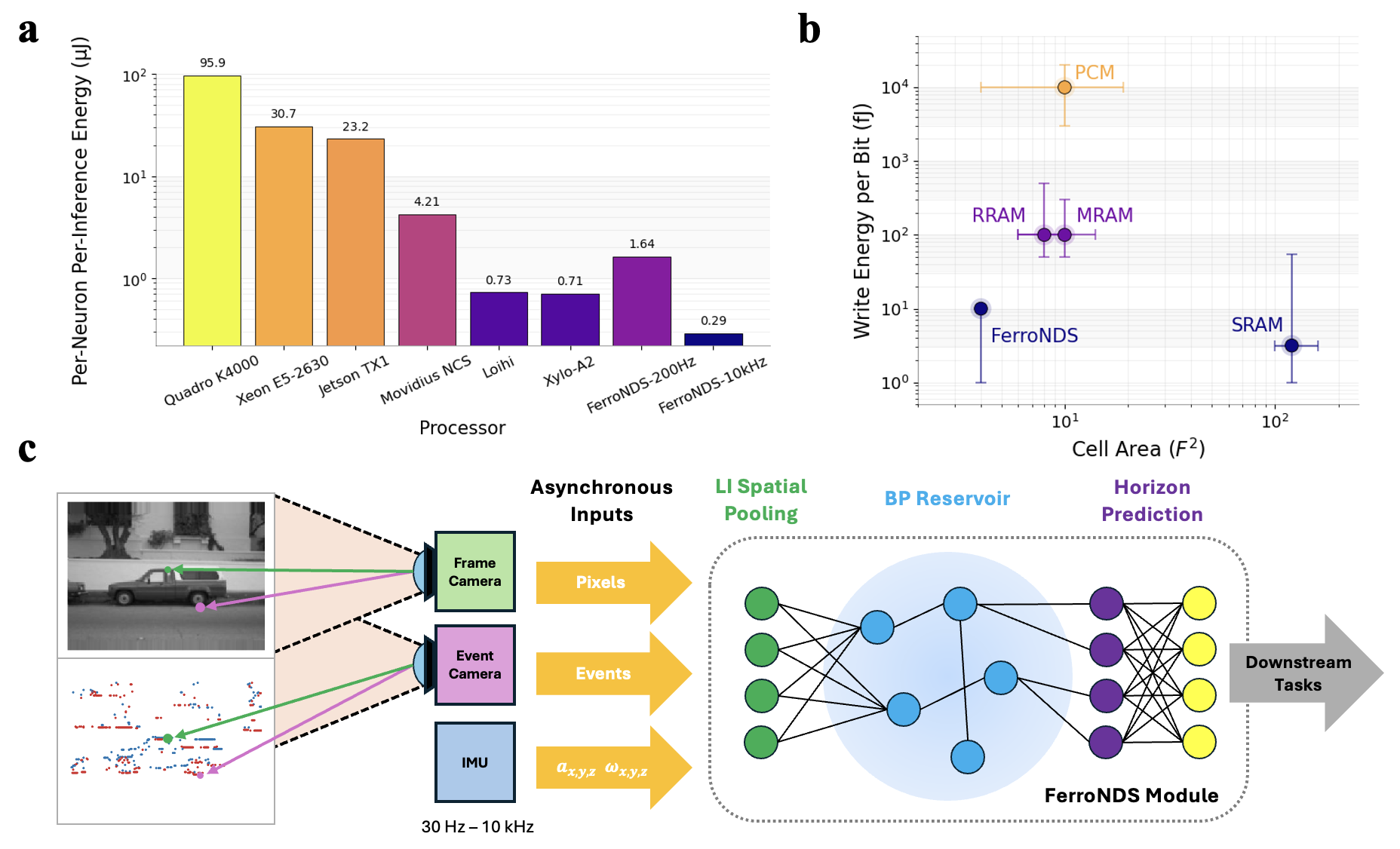}
  \caption{Benchmarking results and target application. \textbf{Benchmarking:} (a) Per-neuron per-inference energy of proposed system at frequency of 200 Hz (FerroNDS-200Hz) and 10 kHz (FerroNDS-10kHz) compared to other single-board computers and neuromorphic processors \cite{bosSubmWNeuromorphicSNN2022}. (b) Cell area versus write energy per bit for five memory technologies. Error bars indicate reported ranges across literature (see Methods). \textbf{Future Work:} (c) Target multi-modal sensing architecture in which frame camera, event camera, and IMU streams (spanning 30 Hz - 10 kHz) are fused through a FerroNDS module that functions as a combined sensor front-end and edge processor for downstream tasks.}
  \label{fig:6}
\end{figure} 

\textbf{Undamped harmonic (UH).} Given that the reservoir and BP/LI approaches use damped oscillation, we considered the case of reduced damping to allow oscillation over much longer time scales. The resulting UH/LI architecture (Algorithm S5) follows recent work which shows that removing the damping parameter captures long-range temporal dependencies by preserving longer oscillations \cite{agrawalSecondOrderSpikingSSMWearables2025}. Integrating undamped dynamics requires care because standard Euler integration can accumulate amplitude error (Section S4), so we use an implicit-explicit (IMEX) scheme that preserves amplitude stability \cite{ruschOscillatoryStateSpaceModels2024}. With this discretization, the UH/LI model predicts 500 ms into the future of Mackey-Glass with inference MSE below 0.003 (Fig. \ref{fig:5}(f); Table S1). Mackey-Glass is a chaotic delay-differential equation whose trajectories are notoriously difficult to predict at long horizons; this level of agreement across the 500-ms window suggests that undamped oscillatory dynamics, by preserving state over long time scales, enable FerroNDS to capture long-range dependencies that the damped architectures cannot. The UH/LI architecture points to a direction worth pursuing: extending undamped oscillator dynamics to longer horizons and spatiotemporal tasks \cite{agrawalSecondOrderSpikingSSMWearables2025}.

\subsection{Benchmarking} \label{sec:5}

\textbf{Total energy.} FerroNDS achieves per-neuron per-inference energy of 1.64 $\mu$J at 200 Hz and 0.29 $\mu$J at 10 kHz (Fig. \ref{fig:6}(a); derivation in Methods). Using a feed-forward architecture matching SynSense Xylo-A2, FerroNDS-200Hz sits far below single-board computers and is comparable to spike-based neuromorphic processors Loihi and Xylo-A2 despite being fully analog \cite{bosSubmWNeuromorphicSNN2022}. FerroNDS-10kHz achieves more than a 2$\times$ energy improvement over both Loihi and Xylo-A2. At 200 Hz, FerroNDS supports systems of up to 3000 neural units within a 1 W power budget; at 10 kHz, the faster op-amps reduce this to 345 neural units. For context, a typical Bluetooth low-energy sensor node operates within a total active power budget of 10-50 mW, of which only a fraction is allocated to on-node inference \cite{khalifehMicrocontrollerUnitBasedWireless2022}. A modestly sized FerroNDS-200Hz system of $\mathcal{O}(10^2)$ neural units falls within this budget, positioning analog neuromorphic inference at scales relevant to battery-powered edge nodes.

\textbf{Lower-voltage projection.} To justify the multi-bit FeD design, we compare its synaptic weighting circuit against an alternative based on a binary FeD (Fig. S2(a)). The multi-bit FeD circuit achieves power and area savings at 3-bit precision and above (Fig. S2(b); full analysis in Section S2). A further path to improvement is voltage reduction. Existing ferrodiode demonstrations have shown robust operation down to 2 V \cite{kimMultistateUltrathinBackEndofLineCompatible2024} \cite{sarkarMultistateFerroelectricDiodes2024} \cite{sarkarCanFerroelectricDiode2025}. At 2 V operation, the input and output scale/bias op-amps become unnecessary, leaving only a sense amp and log-amp. This would reduce synaptic circuit energy by 4.5$\times$ (to 3.4 $\mu$W at 200 Hz) and 20$\times$ (to 6.8 $\mu$W at 10 kHz), yielding an order-of-magnitude improvement over Loihi and Xylo-A2.

\textbf{Area and write energy.} Fig. \ref{fig:6}(b) compares memory cell area (as a function of squared feature size $F^2$) and write energy per bit (fJ) of FerroNDS against SRAM, PCM, RRAM, and MRAM. We find that such a FerroNDS has a 25-40$\times$ area reduction over SRAM ($4\,F^2$ vs $>100\,F^2$) and a 1.5-5$\times$ improvement over PCM, RRAM, and MRAM. In write energy, FerroNDS ($<10$ fJ/bit) is comparable to SRAM (1-10 fJ/bit) while being two to three orders of magnitude lower than PCM ($\sim$10 pJ/bit) and one order of magnitude lower than RRAM and MRAM ($\sim$100 fJ/bit). The combination of SRAM-competitive write energy at a fraction of the cell area is notable. Write-energy improvements matter for any training regime with frequent weight updates, such as Hebbian plasticity. The compact area footprint and low write energy also position FerroNDS favorably for integration with other analog circuitry on a single die.

\textbf{Latency.} A worst-case estimate of the time to first causal output in the cascaded synapse-to-soma path is 3.18 ms per layer at 200 Hz and 63.87 $\mu$s per layer at 10 kHz (derivation in Methods). Each layer in an $L$-layer network therefore operates in real time, since per-layer latency is less than the input signal rate and each layer processes one sample before the next arrives. For a 10-layer network at 10 kHz, total cascaded latency is roughly 0.64 ms, well within real-time constraints for edge sensing applications.

\section{Discussion} \label{sec:6}

Modern inference accelerators are built around massively-parallel dense linear algebra, a template well-suited to machine learning workloads composed of large matrix operations but not the only shape a learning system can take. We designed FerroNDS to be orthogonal to this template, choosing asynchronous over synchronous, analog over digital, recurrent over feedforward, and targeting small-scale, real-time inference at up to 10 kHz instead of offline workloads. The central architectural insight is that oscillator dynamical systems project temporal inputs into a truncated Fourier-like basis, so that a chirp becomes an affine function and an amplitude-modulated sine reduces to a sparse set of spectral components. FerroNDS thus converts streaming temporal signals into spectral features through neural adaptive filtering and performs autoregressive prediction in that feature space. This places the system's strengths in the opposite regime from dense linear algebra accelerators: sparse, continuous-time dynamics rather than dense, synchronous matrix operations.

A speculative target application is illustrated in Fig. \ref{fig:6}(c). We chose to omit spiking representation to test whether a neuromorphic system communicating through continuous membrane potentials would be feasible, accepting higher per-neuron power as the cost. The resulting per-neuron per-inference energy of 0.29 $\mu$J is lower than Loihi's 0.73 $\mu$J, suggesting that analog, rather than spike-based, computing is a viable design point for small-scale, real-time neural dynamical systems. Two paths to further power reduction remain: software pre-distortion that undoes the FeD's exponential transfer characteristics (potentially eliminating the log-amp) and stable multi-bit operation at 2 V. We pursued the hardware-based FeD linearization rather than the software path, accepting additional peripheral power and area, because future integration of FeD arrays will likely require both hardware conditioning and software calibration for robust synaptic weighting.

In summary, FerroNDS is the first end-to-end FeD-based neuromorphic framework and a generic recipe that combines asynchronous analog components, multi-bit compute-in-memory, and dynamical-systems computation. The synaptic weighting circuit, state-selection algorithm, and ideal device model presented here are device-agnostic and can accommodate any multi-bit eNVM. The integrator and oscillator primitives together implement short-time Fourier transform and 500-ms horizon prediction on periodic, quasi-periodic, and chaotic signals at per-neuron per-inference energy of 1.64 $\mu$J (200 Hz) and 0.29 $\mu$J (10 kHz), 25-40$\times$ area reduction over SRAM, and per-layer latency of 3.18 ms (200 Hz) and 63.87 $\mu$s (10 kHz). Future work will validate FerroNDS on recorded event-based and inertial datasets, assess noise resilience of the hardware-software system, and explore tighter co-integration with analog sensor front-ends.

\section{Methods} \label{sec:7}

\subsection{Ferrodiode device model and state selection}

The ferrodiode I-V characteristics used in this work were obtained from previously published measurements of a 20-nm AlScN film with 5-nm AlO$_x$ interlayer, provided in full by the lead authors \cite{kimMultistateUltrathinBackEndofLineCompatible2024}. Each of the 32 device states was fit to an exponential function $I = G_i \exp(A_i V)$ using a two-stage procedure: \texttt{scipy.stats.linregress} on the semi-log transformation provided an initial estimate, which was then refined using \texttt{scipy.optimize.curve\_fit}. Goodness of fit was evaluated via $R^2$ computed directly on the log-current data.

The state selection algorithm identifies a subset of states satisfying four constraints: (1) sufficiently large output current to avoid noise-limited operation, (2) high-quality exponential fits within the selected voltage window, (3) linear spacing of the $A_i$ coefficients across states, and (4) minimal coefficient of variation in the $G_i$ coefficients. For our FeD, a search over the 32 available states identified 8 states in the [6.45, 7.45] V voltage window satisfying all four constraints: output current $\geq 0.5$ nA, exponential fits with $R^2 \geq 0.998$, $A_i$ terms linearly spaced with $R^2 = 0.987$, and $G_i$ terms with coefficient of variation of 0.37 (Section S1, Fig. S1).

\subsection{Circuit simulation and total energy estimate}

All circuit simulations were performed in LTspice using SPICE macromodels of the Analog Devices LT6003 and LTC2068 operational amplifiers. Power and energy calculations were derived from op-amp quiescent current ($I_q$) and supply voltage ($V_{dd}$) specifications: 850 nA at 1.6 V for LT6003 (200 Hz operation) and 7.5 µA at 1.7 V for LTC2068 (10 kHz operation). Per-synapse power was computed as $I_q \cdot (8 \text{ V} + 3 \cdot 3.3 \text{ V})$, accounting for the one high-voltage and three 3.3 V op-amps in each synaptic weighting circuit. Per-neuron power was computed as $I_q \cdot V_{dd}$ for the oscillator op-amp, which dominates per-neuron energy. Per-neuron per-inference energy at a given inference rate $f$ is given by:
\begin{align}
    \frac{P_{\text{neuron}} + P_{\text{synapse}}}{f \cdot N_{\text{neurons}}}
\end{align}
Where $N_{\text{neurons}}$ is the total number of neurons in the system. A representative calculation used 76 neurons and 1632 synapses based on SynSense Xylo-A2, yielding 1.64 $\mu$J at 200 Hz and 0.29 $\mu$J at 10 kHz \cite{bosSubmWNeuromorphicSNN2022}. Per-layer latency was derived from the effective RC time constant of each op-amp:
\begin{align}
    \tau \approx \frac{1}{2\pi f_{-3\text{dB}}}
\end{align}
Which is then multiplied by the number of cascaded op-amp stages per layer (four).

\subsection{Horizon prediction setup}

All horizon prediction tasks used 10-s signals sampled at 1 kHz ($\Delta t = 1$ ms), yielding 10,000 samples per signal. The first 5 s (5,000 samples) were used for training and the remaining 5 s for inference, creating disjoint train/test segments. For each sample index $k$ in the training or inference window, the model received $x[k]$ as input and was trained or evaluated on predicting $x[k+H]$, where $H = 500$ corresponds to a 500-ms horizon. Context size $P = 1$ was used for all architectures except UH/LI on Mackey-Glass, which used $P = 5$. For chirp, envelope-modulated sine, Mackey-Glass, and composite signals, the inference segment contains spectral or dynamical content not present in the training segment, making these tasks effectively out-of-distribution.

Seven signal classes were evaluated. Noisy sine and noisy square waves were generated at 2 Hz with unit amplitude and additive Gaussian noise of standard deviation 0.05. The amplitude-modulated (AM) sine signal uses a carrier frequency uniformly sampled from [2, 6] Hz and a modulation frequency uniformly sampled from [0.3, 1] Hz. The chirp signal uses a linear frequency sweep from an initial frequency in [0.1, 1] Hz to a final frequency in [5, 6] Hz over the full 10-s duration. The envelope-modulated sine uses a carrier frequency uniformly sampled from [2, 5] Hz with a Gaussian envelope:
\begin{align}
    \exp \left(\frac{-(t - T/2)^2}{2(T/5)^2} \right)
\end{align}
Where $T = 10$ s. The composite signal consists of three equal-duration segments at 2 Hz with unit amplitude: a sine wave for the first 3.33 s, a square wave for the second 3.33 s, and a sawtooth wave for the third 3.33 s; training uses the sine and square segments, while inference requires predicting the unseen sawtooth segment. The Mackey-Glass signal was generated by integrating the delay differential equation:
\begin{align}
    \frac{dx}{dt} = \frac{\beta x(t - \tau)}{1 + x(t - \tau)^n} - \gamma x(t)
\end{align}
With $\beta = 0.2$, $\gamma = 0.1$, $n = 10$, $\tau = 17$, and a total trajectory length of $N = 6000$ steps with $\Delta t = 1$, using a fourth-order Runge-Kutta (RK4) integration scheme. In this canonical chaotic regime ($\lambda_{max} \approx 0.01$), the Lyapunov time is approximately 100 time units, setting the natural forecastability horizon. Training and inference segments were taken as indices [201, 3200] and [5001, 5500], respectively.

Evaluation metrics are mean squared error (MSE) and mean absolute error (MAE), both computed per trajectory and averaged across trajectories. Inference was run sample-by-sample.

\subsection{Network architectures and training}

All networks were trained with the Adam optimizer, MSE loss, and no regularization or gradient clipping. 

The feed-forward BP/LI architecture consists of a BP oscillator layer with 64 units with recurrent connections among oscillators, followed by an LI layer with 64 units, followed by a linear readout (Algorithm S3). All parameters are trained end-to-end. The BP layer's feed-forward weights are initialized using Xavier uniform initialization; the adaptive oscillation frequencies $\omega$ are initialized uniformly in [10, 50] Hz; adaptive bias offset terms are initialized uniformly in [1, 6]. The LI layer's feed-forward weights use Xavier uniform initialization, with adaptive time constants $\tau$ initialized from $\mathcal{N}(20, 5)$. Training uses learning rate 0.001, batch size 64, for 50 epochs.

The reservoir-style BP architecture uses a fully-connected pool of 128 BP oscillators with a linear readout. Both internal weights and readout are learned end-to-end. Filter bank oscillators are initialized with resonant frequencies linearly spaced in [1, 64] Hz. Recurrent weights are initialized using \texttt{torch.randn}, with the output gain initialized to ones and the gate initialized via \texttt{nn.Linear}. Training uses learning rate 0.01, batch size 1, for 25 epochs.

The UH/LI architecture follows the structure in Algorithm S5, composed of stacked UH and LI blocks with gated linear units (GLU) and GELU activations. The UH layer uses the IMEX discretization described in Section S4. UH feed-forward weights, learnable projection matrices $C$ and $D$, and GLU weights use default \texttt{nn.Linear} initialization; adaptive frequencies $\omega$ are initialized from a standard normal distribution. LI layers follow the same initialization as in BP/LI. Training uses learning rate 0.001, batch size 64, for 25 epochs.

The MLP baseline is a two-layer ReLU network that receives a single-sample context ($P = 1$) and predicts $x[k+H]$. Feed-forward layers use default \texttt{nn.Linear} initialization (standard normal for weights, zero for biases). Training uses learning rate 0.001, batch size 64, for 50 epochs. 

Basic modules and parameters in \texttt{cells.py} and \texttt{model.py} adapted from \cite{higuchiBalancedResonateandFireNeurons2024}.

\subsection{Area and write-energy benchmarking}

Memory cell area was computed as a multiple of squared feature size $F^2$ at a technology-node-normalized reference, using published values for all five memory technologies \cite{sarkarCanFerroelectricDiode2025} \cite{huDemonstrationHighlyScaled2025}. Write energy values are taken from published measurements of commercial and research-grade memory cells for FeD \cite{sarkarCanFerroelectricDiode2025}, SRAM \cite{wang10FJEnergy2023} \cite{lee175fJBitEnergyEfficient2017} \cite{xiaoBiDirectionalOperandControllableInMemory2024} \cite{clerc032V55fJBit2012}, RRAM \cite{zangenehDesignOptimizationNonvolatile2014}, MRAM \cite{caiHighPerformanceMRAM2017} \cite{patelReducingSwitchingLatency2016}, and PCM \cite{sternUncoveringPhaseChange2021} \cite{leeArchitectingPhaseChange2009}.

\subsection{Software}

Network training and evaluation were performed in Python 3.10 using PyTorch 2.7.0, NumPy 1.23.5, and SciPy 1.10.0. Circuit simulations used LTspice. All code is available at the link given in the Code availability statement.

\section*{Data availability}

The datasets generated and analyzed during this study consist of synthetic signals whose exact parameters and generation procedures are fully specified within the Methods section of this article. Ferrodiode I-V characterization data used to fit the device model were provided by the authors of \cite{kimMultistateUltrathinBackEndofLineCompatible2024} and remain available from the corresponding author upon reasonable request.

\section*{Code availability}
Source code for experiments is publicly available at \url{https://github.com/Keshava-Katti/ferronds}.

\section*{Acknowledgements}

K.K. acknowledges support from the National Science Foundation (NSF) Graduate Research Fellowship Program (GRFP), Fellow ID: 2022338725. Any opinions, findings, and conclusions or recommendations expressed in this material are those of the author(s) and do not necessarily reflect the views of the NSF.

\section*{Author contributions}

K.K., P.C., and D.J. conceived the idea. K.K. conceptualized and designed the FerroNDS, as well as performed the neural dynamical system derivations and simulations. K.K. and A.S. co-designed the FeD, integrator, and oscillator circuits. A.S. performed the SPICE simulations. All authors contributed to the writing of the manuscript.

\section*{Competing interests}

The authors declare no competing interests.

\clearpage
\setcounter{section}{0}
\setcounter{figure}{0}
\setcounter{table}{0}
\setcounter{equation}{0}
\setcounter{algorithm}{0}
\renewcommand{\thesection}{S\arabic{section}}
\renewcommand{\thesubsection}{S\arabic{section}.\arabic{subsection}}
\renewcommand{\theequation}{S\arabic{equation}}
\renewcommand{\thefigure}{S\arabic{figure}}
\renewcommand{\thetable}{S\arabic{table}}
\renewcommand{\thealgorithm}{S\arabic{algorithm}}

\begin{center}
{\Large\bfseries Supplemental Information}\\[0.4em]
{\large Neural dynamical systems on ferroelectric compute-in-memory for real-time forecasting}
\end{center}
\vspace{1.5em}

\section{State selection and bandpass oscillator response to a non-resonant input}

\begin{figure}[H]
  \centering
  \includegraphics[width=1.0\textwidth]{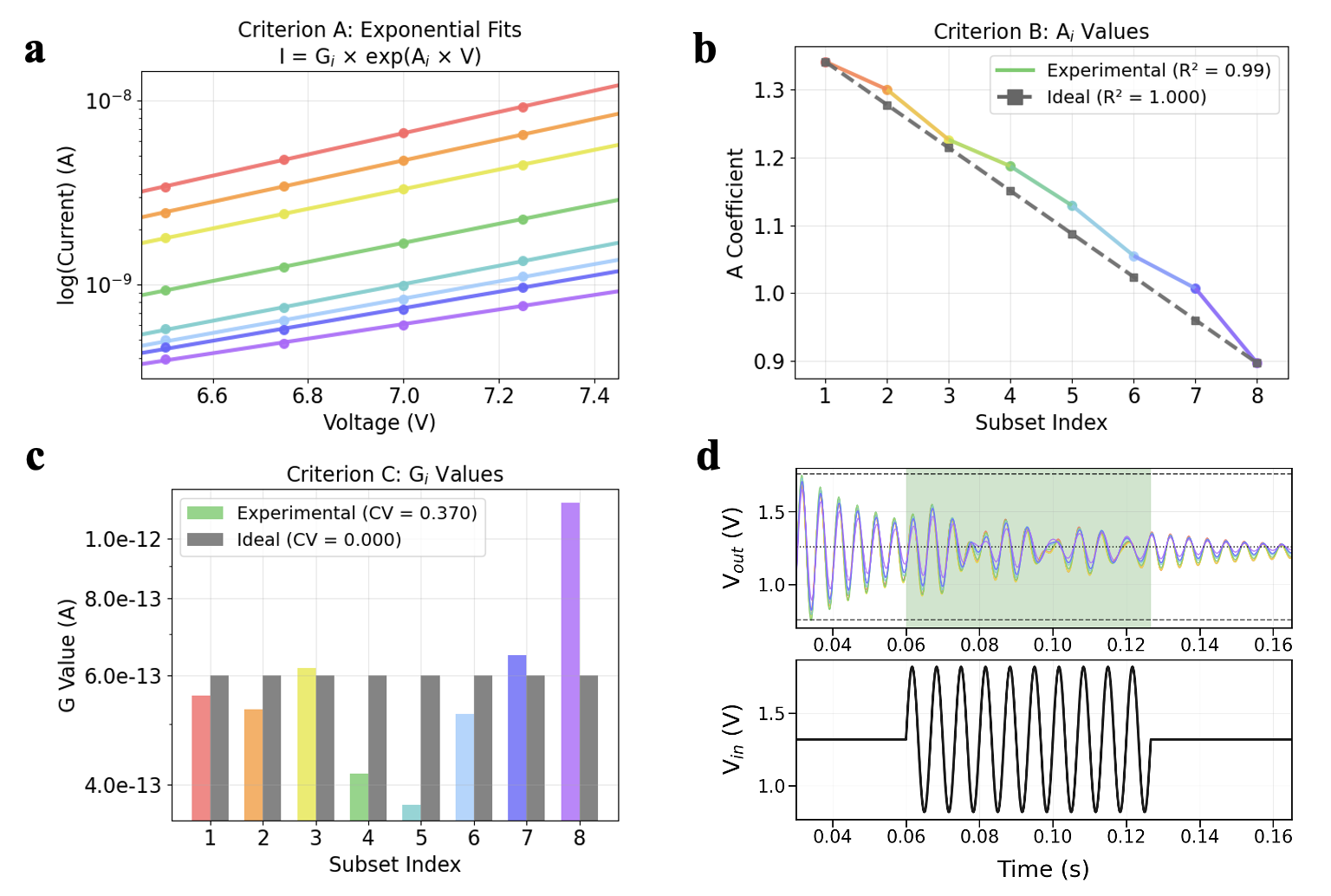}
  \caption{\textbf{State Selection:} Properties of the 8 FeD states selected for the FerroNDS synaptic weighting circuit (Methods). (a) Semi-log I-V curves for the 8 selected states, with exponential fits of $I = G_i \exp(A_i V)$ showing $R^2 \geq 0.998$ across the [6.45, 7.45] V operating window (output current $\geq 0.5$ nA). (b) Extracted $A_i$ coefficients versus state index, showing approximately linear spacing ($R^2 = 0.987$). (c) Extracted $G_i$ coefficients with coefficient of variation 0.37. Together these fits confirm that the 8 selected states satisfy the four constraints for stable multi-bit operation described in Methods. \textbf{Non-Resonant Response:} Response of the BP oscillator circuit (Fig. 3(c)) to a sinusoidal input whose frequency does not match the circuit's resonant frequency. The green region marks the duration of the applied stimulus. Unlike the resonant case shown in Fig. 3(d), the response amplitude remains small during the stimulus and exhibits only subtle damped relaxation to baseline after the stimulus is removed, confirming the circuit's frequency selectivity.}
  \label{fig:supp1}
\end{figure}

\newpage
\section{Binary vs multi-bit synaptic weighting circuit}

\begin{figure}[h]
\centering
\includegraphics[width=\linewidth]{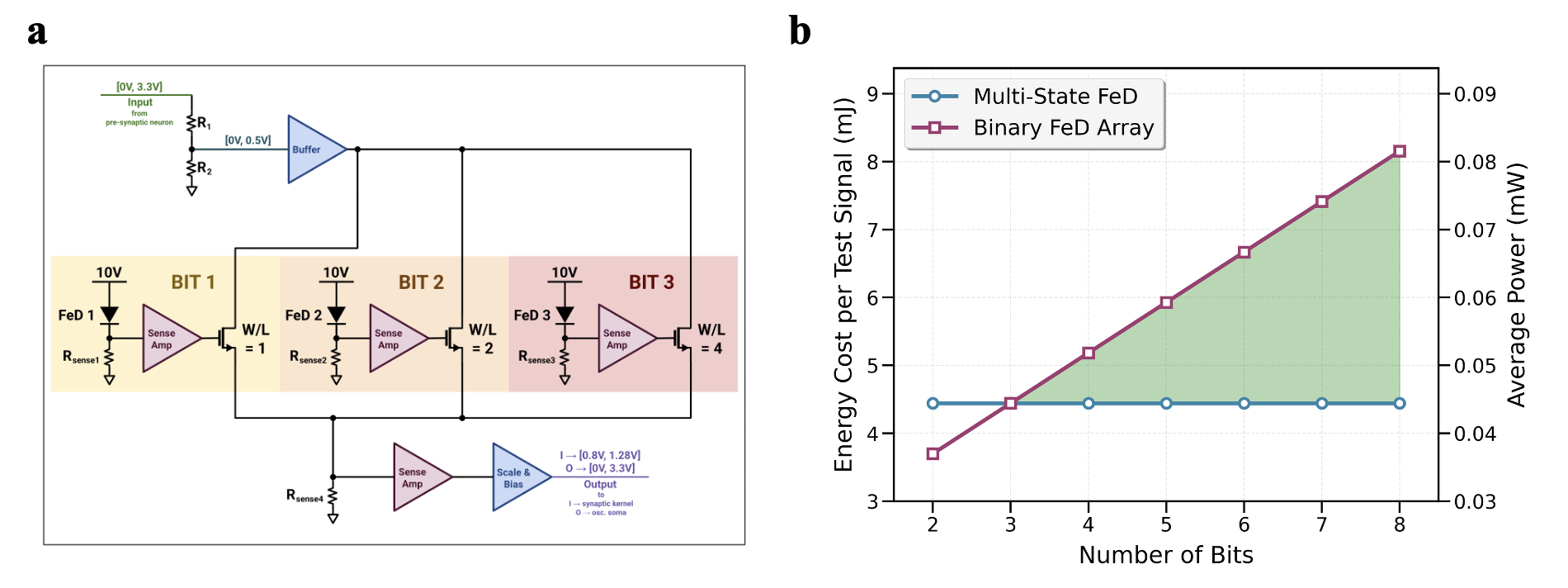}
\caption{Binary vs multi-bit FeD synaptic weighting circuit comparison. (a) Binary FeD weighting circuit using 3 bit cells, each with its own sense amp. This serves as the binary counterpart to the multi-bit FeD circuit shown in Fig. 2(a) of the main paper. (b) Per-circuit energy comparison across bit precisions. The binary circuit's energy scales with the number of bit cells, while the multi-bit circuit's energy is roughly constant. The two circuits reach parity at 3 bits; at higher bit precisions, the multi-bit circuit is increasingly more energy-efficient.}
\label{fig:supp3}
\end{figure}

To validate our hypothesis that a multi-bit FeD-based weighting circuit yields energy savings over its binary counterpart at useful bit precisions, we designed an equivalent binary weighting circuit using 3 FeD bit cells (Fig. \ref{fig:supp3}(a)). While the binary circuit does not require a log-amp, it does require a sense amp for every bit cell. As a result, the binary circuit's energy scales linearly with bit precision, while the multi-bit circuit's energy remains roughly constant. The two circuits reach energy parity at 3 bits; beyond 3 bits, the multi-bit circuit is more efficient (Fig. \ref{fig:supp3}(b)). At 8 bits, the binary circuit consumes roughly 2$\times$ the energy of the multi-bit equivalent.

\newpage
\section{Derivation of integrator and oscillator equation}

Two of the simplest models in computational neuroscience are the leaky integrate-and-fire (LIF) neuron and resonate-and-fire (RF) neuron, both characterized by linear differential equations, a hard firing threshold and reset, and a unique stable equilibrium at rest \cite{izhikevichDynamicalSystemsNeuroscience2007}. The LIF neuron continuously encodes the strength of an input into the frequency of spiking. Many such neurons exist in the cortex and can be characterized by temporal integration of an incoming spike train, with higher input rates eliciting stronger responses \cite{izhikevichResonateandfireNeurons2001}. The RF neuron, by contrast, exhibits frequency preference via damped oscillations. Damped or fast subthreshold oscillations have been observed in many biological neurons, including almost all those described by the Hodgkin-Huxley model \cite{izhikevichResonateandfireNeurons2001}. These neurons prefer inputs with a resonant frequency $\omega$ that matches the frequency of their subthreshold oscillations. Section S3.1 derives the discrete leaky integrator from the LIF equation, and Section S3.2 derives the discrete bandpass oscillator from the RF equation.

Section S3.1 shows how the discrete leaky integrator (LI) equation can be derived from a continuous integrator, with the leaky integrate-and-fire (LIF) equation emerging en route. The LIF model is given by:
\begin{align}
    \dot{y}_i = a_i + b_i y_i + \sum_{j=1}^n s_{ij} \delta (t - t_j^*) \label{3.1}
\end{align}
In this equation, $y_i(t)$ represents the membrane potential of neuron $i$, which evolves continuously in time except at discrete instants when presynaptic spikes arrive or when neuron $i$ itself fires. When $y_i$ crosses some threshold $y_{th}$, neuron $i$ emits a spike and $y_i$ is reset to some baseline $y_{reset}$. The intrinsic dynamics $a_i + b_i y_i$ involve a constant bias current $a_i$ supplied to neuron $i$, as well as a leak term $b_i < 0$ that sets the time constant $\tau_i = -1 / b_i$ that corresponds to exponential relaxation of $y_i$ toward steady-state $y_i^* = -a_i / b_i$. The synaptic input term represents the sum of instantaneous input currents arriving from presynaptic neuron $j$ at time $t_j$. The term $s_{ij}$ is the synaptic weight from neuron $j$ to neuron $i$, which is excitatory if $s_{ij} > 0$ and inhibitory if $s_{ij} < 0$. The Dirac delta function $\delta(t - t_j^*)$ injects a discrete impulse at time $t_j^*$, which represents the nearest moment that the $j$th neuron fires. 

Similarly, Section S3.2 goes from a continuous simple harmonic oscillator to the resonate-and-fire (RF) equation to a discrete bandpass (BP) oscillator. The RF model can be written as:
\begin{align}
    \dot{z}_i = (b_i + \text{i}\omega_i)z_i + \sum_{j=1}^n c_{ij} \delta (t - t_j^*) \label{3.2}
\end{align}
The interpretation of (\ref{3.2}) is slightly more abstract, with $z_i \in \C$ describing the state of the $i$th neuron. If we write $z_i = p_i + \text{i} q_i$, then the real part $p_i$ is a current-like variable while the imaginary part $q_i$ is a voltage-like variable. The internal parameter $b_i + \text{i} \omega_i$ contains the rate of attraction $b_i < 0$ to steady-state and frequency of oscillation $\omega_i > 0$. The term $c_{ij}$ still represents a synaptic weight which could be a complex-valued constant. The RF model also has a threshold $z_{th}$, after which the internal state relaxes via damped oscillations back down to the rest state $z = 0$. 

\subsection{Discrete leaky integrator}

The ideal integrator equation in continuous-time can be written as:
\begin{align}
    \dot{u} = x \label{6.1}
\end{align}
That is, when the derivative of $u$ equals input $x$, the system perfectly integrates the input over time, with solution given by:
\begin{align}
    u(t) = u(0) + \int_0^t x(\tau) d\tau \label{6.2}
\end{align}
The problem with the integrator system (\ref{6.1}) with solution (\ref{6.2}) is that it has an infinite memory, i.e., past inputs accumulate indefinitely. A result is that the system is marginally stable; any constant bias in $x$ could cause unbounded drift. To help alleviate this stability issue, a variant known as a LI is often used. Such a system still integrates the input but ``forgets'' past inputs with exponential decay and is given as follows:
\begin{align}
    \dot{u} = -\frac{1}{\tau} u + x \label{6.3}
\end{align}
The solution to (\ref{6.3}) is given by:
\begin{align}
    u(t) = u(0) e^{-t/\tau} + \int_0^t e^{-(t - \tau') / \tau} x(\tau') d\tau' \label{6.4}
\end{align}
Which now has a finite memory with time constant $\tau$ and is asymptotically stable. Setting $\tau$ to be smaller yields faster exponential decay/forgetting, i.e., a shorter memory into the past. Often rather than just passing in $x$, we choose to pass in a term $\phi(x)$ that captures the synaptic weighting and an optional nonlinearity, which yields the following alternative form of the LI system given by:
\begin{align}
    \dot{u} = -\frac{1}{\tau} u + \phi(x) \label{6.5}
\end{align}
Where we typically set $\phi(x) = Wx$ for some synaptic weighting matrix $W$. Looking at (\ref{6.5}) helps us understand why (\ref{3.1}) is called a LIF neuron. Indeed if we make the following variable replacements and constant/function definitions:
\begin{align}
    \quad y_i \leftarrow u \quad \quad t_j^* \leftarrow x \quad \quad a_i = 0 \quad \quad b_i = -\frac{1}{\tau} \quad \quad \phi(x) = \sum_{j=1}^n s_{ij} \delta(t - x) \label{6.6}
\end{align}
We recover exactly the LI system given in (\ref{6.5}) from the LIF neuron model in (\ref{3.1}). Now notice that if we rewrite (\ref{6.5}) such that the input and leak share the same time constant $\tau$, yielding the following system:
\begin{align}
    \dot{u} = \frac{\phi(x) - u}{\tau} \label{6.7}
\end{align}
We recover the standard first-order LI given in (2.4) by setting $u \leftarrow y$ and $\phi(\cdot)$ to be the identity function. In order to model the system given by (\ref{6.7}), we need to discretize it as follows. We choose a fixed time step $\Delta t$ such that we can discretize at $t_k = k \Delta t$ for $k \in \mathbb{N}$. Furthermore, we define $u_k := u(t_k)$. On each interval $[t_{k-1}, t_k]$, we assume $x(t)$ is constant and equal to its value at the start of the interval, i.e., $\phi(x(t)) \approx \phi(x_{k-1})$ for $t \in [t_{k-1}, t_k]$. Therefore, we have that:
\begin{align}
    \dot{u}(t) = \frac{\phi(x_{k-1}) - u(t)}{\tau}, \quad \quad t \in [t_{k-1}, t_k] \label{6.8}
\end{align}
This is a linear ODE with constant coefficients. The general solution on this interval is:
\begin{align}
    u(t) = Ce^{-t/\tau} + \phi(x_{k-1}) \label{6.9}
\end{align}
By imposing the initial condition at $t = t_{k-1}$, we get that:
\begin{align}
    u_{k-1} = Ce^{t_{k-1} / \tau} + \phi(x_{k-1}) &\implies C = (u_{k-1} - \phi(x_{k-1})) e^{t_{k-1} / \tau} + \phi(x_{k-1}) \nonumber \\
    &\implies u(t) = (u_{k-1} - \phi(x_{k-1})) e^{-(t-t_{k-1}) / \tau} + \phi(x_{k-1}) \label{6.10}
\end{align}
At the end of the interval, we have that $t_k = t_{k-1} + \Delta t$, which gives us that:
\begin{align}
    u_k &= (u_{k-1} - \phi(x_{k-1})) e^{-\Delta t / \tau} + \phi(x_{k-1}) \nonumber \\
    &= e^{-\Delta t / \tau} u_{k-1} - \phi(x_{k-1}) e^{-\Delta t / \tau} + \phi(x_{k-1}) \nonumber \\
    &= e^{-\Delta t / \tau} u_{k-1} - (1 - e^{-\Delta t / \tau}) \phi(x_{k-1}) \label{6.11}
\end{align}
By setting $\alpha := \sigma(e^{-\Delta t / \tau})$, where the sigmoid function $\sigma$ is optionally used to ensure $0 < \alpha < 1$, as well as $\Delta t > 0$ and $\tau > 0$, we get that:
\begin{align}
    u_k = \alpha u_{k-1} + (1 - \alpha) \phi(x_{k-1}) \label{6.12}
\end{align}

\subsection{Discrete bandpass oscillator}

The standard simple harmonic oscillator (SHO) in continuous-time can be written as:
\begin{align}
    \ddot{v} + \omega^2 v = 0 \iff \begin{cases} \dot{u} = -\omega^2 v \\ \dot{v} = u \end{cases} \label{6.13a}
\end{align}
Such an SHO has neither damping nor forcing. We can add a force (i.e., input) $I(t)$ and a damping coefficient $b$ as follows:
\begin{align}
    \ddot{v} + 2 b \dot{v} + \omega^2 v = I(t) \iff \begin{cases} \dot{u} = -2b u -\omega^2 v + I(t) \\ \dot{v} = u \end{cases} \label{6.13b}
\end{align}
To get a coupled first-order form of a damped and forced harmonic oscillator, where $v(t)$ is the oscillator's displacement (analogous to membrane potential), $u(t) = \dot{v}(t)$ is the velocity or current-like variable, $b > 0$ sets the damping coefficient, and $I(t)$ is the external forcing (or input current). Notice that we can exactly recover (\ref{6.13b}) from our BP oscillator equation given by (2.6) by making the following variable replacements and constant definitions:
\begin{align} 
    z_1 \leftarrow u \quad \quad z_2 \leftarrow v \quad \quad b = \xi \omega \label{6.14}
\end{align}
Next, we want to consider the relationship between (\ref{6.13b}) and the resonate-and-fire neurons given by (\ref{3.2}). With a simple set of substitutions, (\ref{3.2}) can be written as:
\begin{align}
    \dot{u} = (b + \text{i}\omega) u + I(t) \iff \begin{cases} \dot{x} = bx -\omega^2 y + I(t) \\ \dot{y} = \omega x + by \end{cases} \label{6.15}
\end{align}
Now let us differentiate the first equation to get:
\begin{align}
    \ddot{x} = b \dot{x} - \omega \dot{y} + \dot{I}(t) \label{6.16}
\end{align}
Then we substitute the second equation $\dot{y} = \omega x + by$ to get:
\begin{align}
    \ddot{x} = b \dot{x} - \omega (\omega x + by) + \dot{I}(t) \label{6.17}
\end{align}
By substituting a rearranged version of the first equation $y = (bx - \dot{x} + I(t)) / \omega$ into (\ref{6.17}), we get that:
\begin{align}
    \ddot{x} = b\dot{x} - \omega^2 x - b(bx - \dot{x} + I(t)) + \dot{I}(t) \implies \ddot{x} + 2b\dot{x} + (\omega^2 - b^2) x = \dot{I}(t) - bI(t) \label{6.18}
\end{align}
If we neglect the small $b^2$ term (for weak damping), then we have that:
\begin{align}
    \ddot{x} + 2b\dot{x} + \omega^2 x \approx \dot{I}(t) - bI(t) \label{6.19}
\end{align}
By comparing (\ref{6.19}) with (\ref{6.13b}), we see that the damped and forced harmonic oscillator is essentially a smoothed (i.e., low-pass) version of the resonate-and-fire neuron. That is, it absorbs the $\dot{I}(t)$ term into the forcing and eliminates the explicit rotational structure. To summarize, our BP oscillator, which is a specific instance of (\ref{6.13b}), is a real second-order form of the resonate-and-fire neuron given by (\ref{6.15}) after eliminating $y$. Just as we did with the LI in Section S3.1, we need to discretize the BP oscillator. As mentioned in Section 2.2, we can do so by using Euler integration to get the following discrete version:
\begin{align}
    u_k &= u_{k-1} + \Delta t \left(-2b u_{k-1} - \omega^2 v_{k-1} + I_k\right) \nonumber \\
    v_k &= v_{k-1} + \Delta t \left(u_{k-1}\right) \label{6.20}
\end{align}

\newpage
\section{Model architecture}

\textbf{BP/LI architecture.} The BP oscillator captures behavior in an Andronov-Hopf oscillator when $\alpha < 0$ (i.e., stable focus). The subroutine for how we implemented the BP oscillator, LI module, and the combined feed-forward BP/LI architecture is shown in Subroutine \ref{alg:1}, Subroutine \ref{alg:2}, and Algorithm \ref{alg:3}, respectively.

\floatname{algorithm}{Subroutine}
\begin{algorithm}
\caption{Bandpass (BP) Oscillator} \label{alg:1}
\begin{algorithmic}
\Function{BPCell}{$x$}
    \State $x \gets \texttt{Linear}(x)$ 
    \State $b \gets 0.005 \cdot \omega^2 + b_{\text{offset}} \cdot \Delta t - \omega^2 v \cdot \Delta t$ \\

    \State $v \gets v + u \cdot \Delta t$ \Comment{Euler update for BP oscillator}
    \State $u \gets u + x \cdot \Delta t - 2bu$ \\
    
    \State $u \gets \texttt{Normalize}(u)$
    \State $u \gets G \cdot u$ \Comment{Learnable gain term $G$}
\EndFunction
\end{algorithmic}
\end{algorithm}

\begin{algorithm}
\caption{Leaky Integrator (LI)} \label{alg:2}
\begin{algorithmic}
\Function{LICell}{$x$}
    \State $x \gets \texttt{Linear}(x)$
    \State $\alpha \gets \exp(-1 / \tau_{mem})$
    \State $u \gets u \cdot \alpha + x (1 - \alpha)$ \Comment{Euler update for LI}
\EndFunction
\end{algorithmic}
\end{algorithm}

\floatname{algorithm}{Algorithm}
\begin{algorithm}
\caption{BP/LI NDS} \label{alg:3}
\begin{algorithmic}
\Require Input sequence $x$
\Ensure Sequence length $T$, output sequence $y$
\For{$t = 1, \dots, T$}
    \State $u_t \gets \texttt{BPCell}(x_t)$
    \State $y_t \gets \texttt{LICell}(u_t)$
\EndFor \\
\Return{$y$}
\end{algorithmic}
\end{algorithm}

\textbf{UH/LI architecture.} Implicit-explicit (IMEX) discretization methods treat the stiff terms implicitly and the non-stiff terms explicitly, yielding a balanced scheme that enables stable yet undamped oscillations. Further, such schemes preserve the total energy of the system and therefore are well-suited for long-range sequential patterns without introducing artificial dissipation. Specifically, IMEX is able to remain stable with a relatively large step size $\Delta t = 0.05$ s when integrating a simple harmonic oscillator since it behaves like a symplectic Euler method, i.e., it preserves a modified energy and keeps the amplitude essentially constant over long times. RK4 is not symplectic, but it is 4th-order accurate, so for a simple linear oscillator even at $\Delta t = 0.05$ s, its numerical solution is extremely close to the exact sinusoid. In contrast, Euler has eigenvalues outside the unit circle and hence the amplitude grows, while implicit (IM) has eigenvalues strictly inside the unit circle, causing the amplitude to decay. Algorithm \ref{alg:4} follows the approach in \cite{agrawalSecondOrderSpikingSSMWearables2025}, where IMEX discretization gives that the undamped harmonic case is:
\begin{align}
    u_k &= u_{k-1} + \Delta t (-\Omega v_{k-1} + W x_k) \nonumber \\
    v_k &= v_{k-1} + \Delta t u_k \label{4.3}
\end{align}
Where $\Omega$ is a diagonal matrix of oscillation frequencies and $W$ is an input projection matrix. By introducing the matrix inverse $S = (I + \Delta t^2 \Omega)^{-1}$, we get the following update rule:
\begin{align}
    s_k = M^{\text{IM}} s_{k-1} + F_k^{\text{IM}} \label{4.4}
\end{align}
Where:
\begin{align}
    M^{\text{IM}} = \begin{bmatrix}
        S & -S \Delta t \Omega \\
        S \Delta t & S
    \end{bmatrix}, \quad \quad F_k^{\text{IM}} = \begin{bmatrix}
        S \Delta t W x_k \\
        S \Delta t^2 W x_k
    \end{bmatrix} \label{4.5}
\end{align} 
The undamped harmonic oscillator captures behavior in an Andronov-Hopf oscillator when $\alpha > 0$ (i.e., limit cycle). Algorithm \ref{alg:5} combines the ``Oscillatory State-Space Model'' approach in \cite{ruschOscillatoryStateSpaceModels2024} with UH/LI units. 

\floatname{algorithm}{Subroutine}
\begin{algorithm}
\caption{Undamped Harmonic (UH)} \label{alg:4}
\begin{algorithmic}
\Function{UHCell}{$x$}
    \For{$t = 1, \dots, T$}
        \State $x_t \gets \texttt{Linear}(x_t)$
        \State $u_t \gets u_{t-1} + (-\Omega v_{t-1} + x) \Delta t$  \Comment{IMEX update for UH oscillator}
        \State $v_t \gets v_{t-1} + u_t \Delta t$
    \EndFor
\EndFunction
\end{algorithmic}
\end{algorithm}

\floatname{algorithm}{Algorithm}
\begin{algorithm}
\caption{UH/LI NDS} \label{alg:5}
\begin{algorithmic}
\Require Input sequence $x$
\Ensure $L$-blocks, output sequence $y$
\State $x_0 \gets \texttt{Linear}(x)$
\For{$\ell = 1, \dots, L$}
    \State $y_{\ell} \gets \texttt{UHCell}(x_{\ell-1})$
    \State $y_{\ell} \gets \texttt{LICell}(y_{\ell})$
    \State $u_{\ell} \gets C y_{\ell} + D (x_{\ell})$ \Comment{Learnable terms $C, D$}
    \State $u_{\ell} \gets \texttt{GELU}(u_{\ell})$
    \State $x_{\ell} \gets \texttt{GLU}(u_{\ell}) + x_{\ell - 1}$ \Comment{Gated Linear Unit (GLU)}
\EndFor
\State $y \gets \texttt{Linear}(y_L)$ \\
\Return{$y$}
\end{algorithmic}
\end{algorithm}

\newpage
\section{MLP baseline performance on periodic and quasi-periodic signals}

\begin{figure}[H]
\centering
\includegraphics[width=0.78\linewidth]{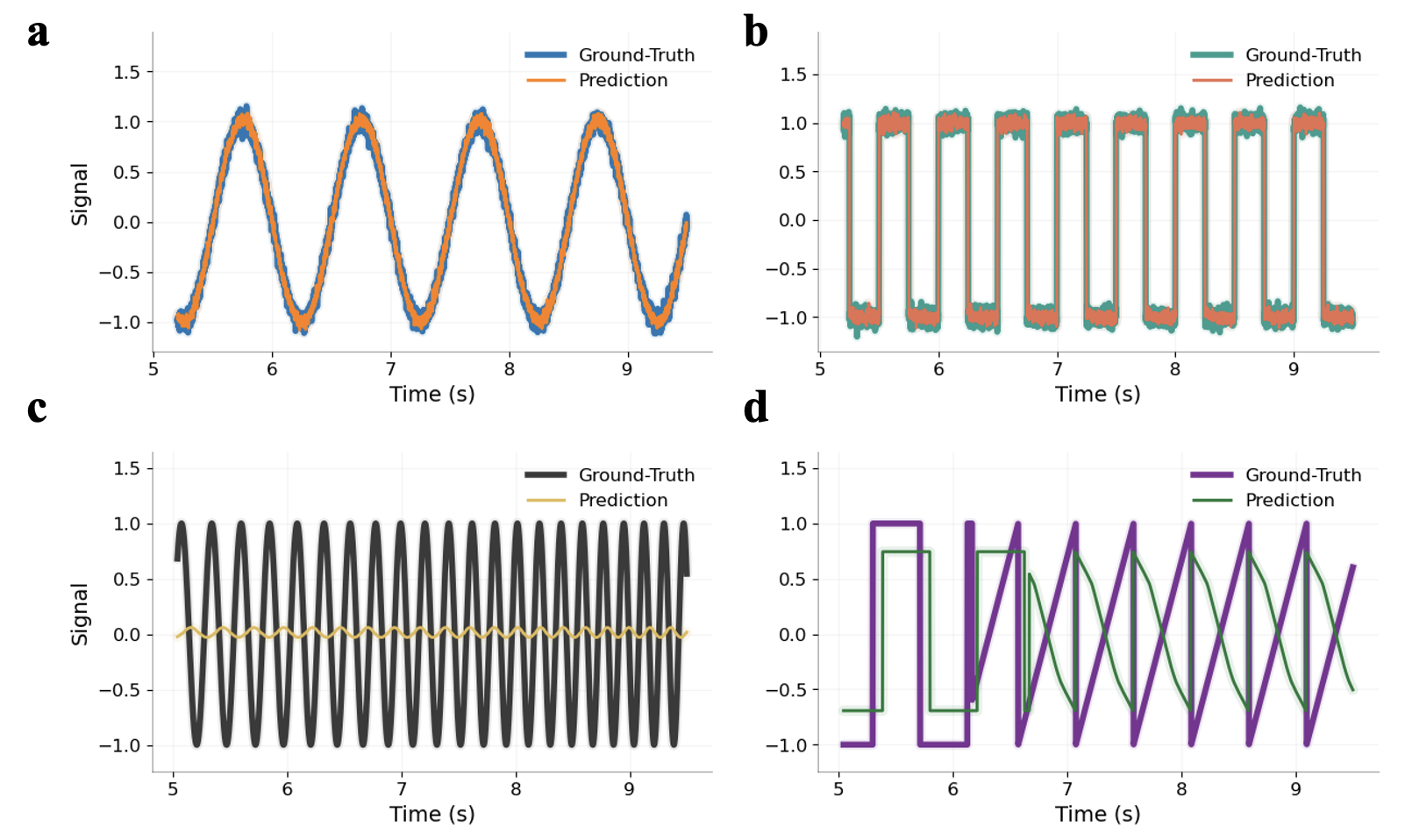}
\caption{Multi-layer perceptron baseline on horizon prediction. Performance of a 2-layer ReLU MLP with single-sample context ($P = 1$) predicting 500 ms into the future, trained and evaluated as described in Methods. (a, b) The MLP predicts highly-structured periodic signals accurately: a 2 Hz sine wave with additive Gaussian noise (a) and a 2 Hz square wave with additive Gaussian noise (b). Both achieve MSE below 0.008 (Table S1). (c, d) The same MLP struggles on quasi-periodic signals: a chirp (c), where training occurs during the slow-moving early portion and inference must predict the faster-moving later portion, and a composite waveform (d), where training uses sine and square segments but inference requires predicting an unseen sawtooth segment. MSE values for these tasks exceed 0.49 (Table S1), indicating that a dense feed-forward architecture does not generalize across distributional shifts in spectral content.}
\label{fig:supp2}
\end{figure}

\newpage
\section{Horizon prediction performance}

\begin{table}[H]
  \centering
  \begin{tabular}{llllcc}
    \toprule
    \textbf{Architecture} & \textbf{Signal Class} & \textbf{Signal} & \textbf{MSE} & \textbf{MAE} \\
    \midrule
    Reservoir BP        & Quasi-periodic & Composite (sine/square/sawtooth) & 0.308 & 0.268 \\
    \midrule
    BP/LI (feed-forward) & Periodic       & AM sine                          & 0.046 & 0.179 \\
    BP/LI (feed-forward) & Quasi-periodic & Chirp                            & 0.319 & 0.444 \\
    BP/LI (feed-forward) & Quasi-periodic & Envelope-modulated sine          & 0.038 & 0.168 \\
    \midrule
    UH/LI               & Chaotic        & Mackey-Glass                     & 0.003 & 0.038 \\
    \midrule
    MLP (baseline)      & Periodic       & Noisy sine                       & 0.004 & 0.051 \\
    MLP (baseline)      & Periodic       & Noisy square                     & 0.008 & 0.049 \\
    MLP (baseline)      & Quasi-periodic & Chirp                            & 0.499 & 0.636 \\
    MLP (baseline)      & Quasi-periodic & Composite (sine/square/sawtooth) & 0.655 & 0.643 \\
    \bottomrule
  \end{tabular}
  \vspace{3mm}
  \caption{Horizon prediction performance across FerroNDS architectures and a multi-layer perceptron (MLP) baseline. All results are for 500-ms horizon prediction on 10-s signals, with training and inference performed on disjoint segments of each signal. For chirp, envelope, Mackey-Glass, and composite signals, the inference split is out-of-distribution relative to training. Errors are reported as mean squared error (MSE) and mean absolute error (MAE).}
  \label{tab:horizon_prediction}
\end{table}

\newpage

\clearpage
\bibliographystyle{unsrt}
\bibliography{references}

\end{document}